\documentclass[twocolumn,letterpaper,amsmath,amssymb,floatfix,aps,superscriptaddress]{revtex4-1}

\usepackage{hyperref}
\usepackage{verbatim}
\usepackage{graphicx}
\usepackage{bm} 
\usepackage{ifpdf} 
\usepackage{dcolumn}  
\usepackage{epsfig}
\usepackage{color}

\newcommand{\Av}[1]{{\mathbf #1}}
\newcommand{\Avg}[1]{{\boldsymbol #1}} 

\def\ln{{\operatorname{ln}}}
\def\det{{\operatorname{det}}}

\def\rmd{{\mathrm{d}}}
\def\rmi{{\mathrm{i}}}
\def\rme{{\mathrm{e}}}

\begin{document}

\title{Asymmetric Coulomb fluids at randomly charged dielectric interfaces: Anti-fragility, overcharging and charge inversion}

\author{Ali Naji}
\thanks{Email: \texttt{a.naji@ipm.ir}}
\affiliation{School of Physics, Institute for Research in Fundamental Sciences (IPM), P.O. Box 19395-5531, Tehran, Iran}

\author{Malihe Ghodrat}
\affiliation{School of Physics, Institute for Research in Fundamental Sciences (IPM), P.O. Box 19395-5531, Tehran, Iran}

\author{Haniyeh Komaie-Moghaddam}
\affiliation{School of Physics, Institute for Research in Fundamental Sciences (IPM), P.O. Box 19395-5531, Tehran, Iran}

\author{Rudolf Podgornik}
\affiliation{Department of Theoretical Physics, J. Stefan Institute, SI-1000 Ljubljana, Slovenia}
\affiliation{Department of Physics, Faculty of Mathematics and Physics, University of Ljubljana, SI-1000 Ljubljana, Slovenia}

\begin{abstract}
We study  the distribution of multivalent counterions next to a dielectric slab, bearing a quenched, random distribution of charges on one of its solution interfaces, with a given mean and variance, both in the absence and in the presence 
of a bathing monovalent salt solution. We use the previously derived approach based on the {\em dressed multivalent-ion theory} that combines aspects of the strong and weak coupling of multivalent and monovalent ions in a single framework. The presence of quenched charge disorder on the charged surface of the dielectric slab is shown to substantially increase the density of multivalent counterions in its vicinity. In the counterion-only model (with no monovalent salt ions), the surface disorder generates  an additional logarithmic attraction potential and thus an algebraically singular counterion density profile at the surface. This behavior persists also in the presence of a monovalent salt bath and results in significant violation of the contact-value theorem, reflecting the {\em anti-fragility} effects of the disorder that drive the system towards a more `ordered' state. In the presence of an interfacial dielectric discontinuity, depleting the counterion layer at the surface, the charge disorder still generates a much enhanced counterion density further away from the surface. Likewise, the charge inversion and/or overcharging of the surface occur more strongly and at smaller bulk concentrations of multivalent counterions when the surface carries quenched charge disorder. Overall, the presence of quenched surface charge disorder leads to sizable effects in the distribution of multivalent counterions in a wide range of realistic parameters and typically within a distance of a few nanometers from the charged surface. 
\end{abstract}


\maketitle

\section{Introduction}

Nature is not perfect and various types of disorder are ubiquitous.  
Disorder often causes large changes in the properties of condensed matter systems as predicted on the basis of naive idealized models that assume perfect regularity. Electron properties in two-dimensional \cite{clarke} and three-dimensional disordered media \cite{pwa,phystoday}, crystalline lattices with structural defects \cite{mermin,kleinert}, spin glasses with random interactions \cite{spinglass_rev,parisi} and systems exhibiting criticality modified by the presence of disorder \cite{weinrib}, are all instances of pronounced disorder effects in the bulk of the materials that can fundamentally change the behavior of idealized model systems. 
Apart from their fundamental importance in modifying the bulk properties, disorder effects at surfaces and interfaces are particularly important in the context of the solid-electrolyte interphases \cite{verma,Hribar}  relevant also for energy generation and storage technologies \cite{storage1,storage2}. The structural disorder in the charge distribution and/or dielectric response spatial profile in the vicinity of the material interfaces couple to long-range electrostatic interactions, leading effectively to long-range disorder effects as well \cite{ali-rudi,rudiali,partial,disorder-PRL,jcp2010,pre2011,epje2012,jcp2012,book,pra2009,pre2010}, that cannot be understood in terms of the usual assumptions of piecewise homogeneous charge distribution and/or dielectric properties, underpinning  so much of colloid science and electrochemistry \cite{Israelachvili,VO,Safranbook,Hunter,andelman-rev,holm}.

The coupling between electrostatic interactions and disorder has been already noted and discussed in other important cases \cite{manne1,manne2,Meyer,Meyer2,klein,klein1,klein2,andelman-disorder,kantor-disorder0,kantor-disorder1,kantor-disorder2,Ben-Yaakov-dis,netz-disorder,netz-disorder2,Lukatsky1,Lukatsky2,Rabin}, including surfactant-coated surfaces \cite{Meyer,Meyer2,klein,klein1,klein2}, random polyelectrolytes and polyampholytes \cite{kantor-disorder0,kantor-disorder1,kantor-disorder2,andelman-disorder}, and contaminant adsorption onto macroscopic surfaces or in amorphous films showing grain structure after being deposited on crystalline substrates \cite{liu}. In all these cases the charge distribution often shows a fundamentally disordered component that often remains unaltered after the assembly or fabrication of the materials, thus exhibiting a frozen, or {\em quenched}, type of disorder  (see, e.g., Refs. \cite{netz-disorder,netz-disorder2,andelman-disorder,disorder-PRL,jcp2010,safran1,safran2,safran3,Olvera0} for examples of surfaces with  {\em annealed} charge distributions that will not be considered in this paper). 
This charge disorder coupled to the long-range electrostatic interactions can then leave its fingerprint also on the interactions between macromolecular surfaces that in their turn can play a fundamental role in the stability of colloidal systems  \cite{ali-rudi,rudiali,partial}.

In fact, this coupling between disorder and Coulomb interactions has been suggested to underly the anomalously long-ranged interactions observed in ultrahigh sensitivity experiments on Casimir-van der Waals interactions between surfaces {\em in vacuo} \cite{kim1,kim2,kim3,kim4,tang}. The intricate experimental details of accurate measurements of these interactions can be properly accounted for only if one considers also the disordered nature of charges on and within the interacting surfaces by invoking the so-called {\em patch effect} \cite{disorder-PRL,jcp2010,pre2011,epje2012,jcp2012,book,pra2009,pre2010,barrett,speake,kim1,kim2,kim3,kim4,tang}, where the disorder stems, for instance, from the adsorption of charged contaminants and/or impurities that can give rise to monopolar random surface charges, and/or the variation of the local crystallographic axes of the exposed surface of a clean polycrystalline sample and the corresponding electron work function that can cause a variation of the local surface potential.  Such random distributions of surface charges can be measured directly by Kelvin force 
microscopy measurements \cite{science11}. 

The salient features of electrostatic interactions themselves, even for homogeneous charge distributions in the absence of any disorder, are however quite involved (see, e.g., Refs. \cite{book,Israelachvili,VO,Safranbook,andelman-rev,holm,Hunter,French-RMP,Shklovs02,Levin02,hoda_review,perspective,Naji_PhysicaA,Edwards,Podgornik89,Podgornik89b,Netz-orland,asim,Netz01,AndrePRL,AndreEPJE} and references therein). It has been recognized some time ago that electrostatic interactions in fact come in several varieties, depending on the strength of electrostatic coupling in the system \cite{Shklovs02,Levin02,hoda_review,perspective,Netz01,AndrePRL,AndreEPJE,Naji_PhysicaA,asim}. In the presence of mobile monovalent counterions, they are standardly described by the Poisson-Boltzmann (PB) theory stemming from the mean-field,  collective description of Coulomb fluids \cite{book,Israelachvili,VO,Hunter,Safranbook,andelman-rev,holm,Podgornik89,Podgornik89b,Netz-orland} that gives rise to pronounced repulsive interactions between like-charged macromolecules (such as polymers, colloids and nano-particles). On the contrary, in the presence of multivalent counterions,  electrostatic interactions exhibit basically a single-particle character and mediate strong attractive interactions between like-charged macromolecules \cite{Shklovs02,Levin02,hoda_review,perspective,Netz01,AndrePRL,AndreEPJE,Naji_PhysicaA,asim}. This attraction led to a new understanding of the theory of electrostatic interactions in colloidal domain based on the {\em strong-coupling (SC) limit}  \cite{Netz01,AndrePRL,AndreEPJE}, devised to describe the equilibrium properties of Coulomb fluids when charges involved become large. In the simple case of a counterion-only system, the transition from the mean-field PB description, dubbed also the {\em weak-coupling (WC) limit},  to the SC limit is governed by a single dimensionless {\em electrostatic coupling parameter}   \cite{Netz01,AndrePRL,AndreEPJE,hoda_review,perspective,Naji_PhysicaA,asim}, being a ratio of the Bjerrum length, which identifies Coulomb interaction between counterions themselves, and the Gouy-Chapman length, which describes electrostatic interaction between the counterions and the charged (macromolecular) surfaces.  The emerging picture of equilibrium properties of Coulomb fluids has thus become much richer than conveyed for many years by the standard DLVO paradigm of colloid science \cite{Israelachvili,VO,Safranbook,Hunter,andelman-rev,holm}.

However, this is still not the complete story. The most relevant case of a Coulomb fluid is in fact not a counterion-only system, but an asymmetric mixture of multivalent ions in a bathing solution of monovalent ions, a particularly relevant situation specifically in the context of bio-macromolecules, where multivalent ions together with the screening properties of the monovalent salt are believed to play a key role in the stability of macromolecular aggregates such as  liquid crystalline mesophases of semiflexible biopolymers \cite{rau-1,rau-2,Angelini03}, or DNA condensates that form in the bulk \cite{Bloom2,Yoshikawa1,Yoshikawa2,Pelta,Pelta2,Plum,Raspaud} or within viruses or virus-like nano-capsids  \cite{Savithri1987,deFrutos2005,Siber}.

In an asymmetric mixture, multivalent counterions and monovalent ions are coupled differently to the macromolecular charges: multivalent ions strongly, while monovalent ions only weakly, as evidenced from their respective electrostatic coupling parameters. Since usually multivalent ions are present at very low concentrations, e.g., around just a few mM, their behavior is expected to be properly described within the virial expansion in  powers of their fugacity (or bulk concentration) \cite{Netz01,SCdressed1,SCdressed2,SCdressed3,perspective,Naji_PhysicaA,hoda_review,asim}. A {\em dressed multivalent-ion theory} then emerges naturally within this context \cite{SCdressed1,SCdressed2,SCdressed3,perspective,leili1,leili2} since the degrees of freedom due to weakly coupled monovalent ions can be traced out from the partition function, leading to an effective formalism based on screened interactions between the remaining dressed multivalent ions and fixed macromolecular charges.  The dressed multivalent-ion theory for complicated asymmetric mixtures of multivalent ions in a bathing solution of monovalent ions can then seamlessly bridge between the standard WC and SC limits \cite{SCdressed1,SCdressed2,SCdressed3,perspective}. 

These baroque features of electrostatic interactions furthermore give their imprint also on the effects of disordered charge distribution along macromolecular interfaces \cite{ali-rudi,rudiali,partial,book,manne1,manne2,Meyer,Meyer2,klein,klein1,klein2,andelman-disorder,kantor-disorder0,kantor-disorder1,kantor-disorder2,Ben-Yaakov-dis,netz-disorder,netz-disorder2,Lukatsky1,Lukatsky2,Rabin,safran1,safran2,safran3,Olvera0}. While on the WC level and for  homogeneous planar systems the quenched disorder effects are nonexistent \cite{ali-rudi}, they can lead to qualitative changes in the stability properties of the system once the dielectric contrast between the solution and macromolecular interfaces (or the inhomogeneous distribution of salt ions) is taken fully into account  \cite{rudiali,Ben-Yaakov-dis,disorder-PRL,jcp2010,pre2011,epje2012,jcp2012,book,pra2009,pre2010}.  
Nevertheless, it is in the SC limit that the coupling between electrostatic interactions and the quenched disorder in the external interfacial charge distributions gives rise to fundamentally novel and unexpected phenomena \cite{ali-rudi,partial}. While studying the interaction between two disordered charged surfaces it was noticed \cite{ali-rudi,partial} that disorder can in fact lead to a lowering of the effective temperature of the system, engendering a distribution of the multivalent counterions between the interacting surfaces that is characterized by less effective entropy. This is intuitively difficult to foresee, as one would perhaps naively assume that thermal and externally imposed charge disorder would somehow enhance one another.  

 In order to properly understand and identify all salient features of the coupling between quenched charge disorder and long-range electrostatic interactions, we now proceed to characterize more closely the consequences of coupling between charge disorder and {\em electrostatically strongly coupled} multivalent counterions immersed in a monovalent salt solution bath. In particular, we will identify the defining feature of this strongly coupled, disordered system as belonging to the {\em anti-fragility} \cite{taleb} exhibited by this system.  In the present context,  anti-fragility simply refers to the fact that an externally imposed, quenched charge disorder, effectively diminishes the intrinsic thermal disorder in the system, forcing its behavior to be more `ordered'. We will show that this behavior stems from the interplay between the {\em translational entropy} of the multivalent counterions and the {\em configurational entropy} due to the averaging over different realizations of the quenched disorder. In the particular example of the counterion-only model (with no monovalent salt ions and no interfacial dielectric discontinuity), we show that multivalent counterions experience an additional logarithmic attraction towards the surface due to the presence of the surface charge disorder in a way that their density profile  exhibits an algebraically singular behavior at the surface with an exponent that depends on the disorder strength (variance). This behavior persists also in the presence of a monovalent salt bath and results in significant violation of the contact-value theorem \cite{contact_value,contact_value1,contact_value2,contact_value3}. 
In the presence of an interfacial dielectric discontinuity, depleting the counterion layer at the surface, the charge disorder still generates a much enhanced counterion density further away from the surface. Likewise, the charge inversion and/or overcharging of the surface are predicted to occur more strongly and at smaller bulk concentrations of multivalent counterions when the surface carries quenched charge disorder. 
 
The organization of the paper is as follows: In Section \ref{sec:model}, we introduce our model and then in Section \ref{sec:formal} present the theoretical formalism that will be used to study the distribution of multivalent counterions next to a randomly charged dielectric interface. We then proceed to present our results in Section \ref{sec:results}, where we discuss the case of counterion-only systems and the effects due to the presence of a bathing salt solution, an interfacial dielectric discontinuity, and also the overcharging and charge-inversion phenomena in the system. We conclude our discussion in Section \ref{sec:discussion}.

\section{The Model}
\label{sec:model}

\begin{figure}[t!]
\begin{center}
	\includegraphics[width=8cm]{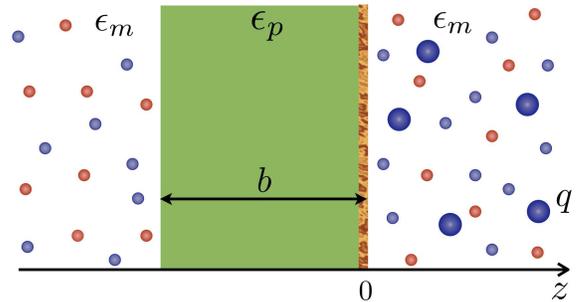}
\caption{(Color online) Schematic view of an infinite, planar dielectric slab of thickness $b$ and dielectric constant $\epsilon_p$ immersed in a solution of dielectric constant $\epsilon_m$, containing a mixture of mono- and multivalent salts. The slab boundary at $z=0$ is randomly charged. The multivalent ions (counterions) with charge valency $q$ are shown by large blue spheres and monovalent salt anions and cations are shown by small orange and blue spheres. The multivalent counterions are in contact only with the charged  surface (right).}
\label{f:schematic}
\end{center}
\end{figure}

Our model is comprised of an infinite, planar dielectric slab of thickness $b$ and dielectric constant $\epsilon_p$, immersed in a bathing ionic solution of dielectric constant $\epsilon_m$ (see Fig. \ref{f:schematic}). The
ionic solution is assumed to consist of a mixture of a monovalent 1:1 salt of bulk concentration $n_0$ as well as of a multivalent $q$:1 salt of bulk concentration $c_0$. The  dielectric slab is assumed to be impermeable to mobile ions and occupy the region $-b\leq z\leq 0$ with its surface normal oriented in the direction of the $z$-axis. The bounding surface of the slab at $z=0$ is assumed to be charged, carrying a {\em quenched} spatial distribution of random monopolar charges, $\rho({\mathbf r})$, while the other surface of the slab is uncharged. The multivalent counterions are assumed to be in contact only with the charged surface (see below). The disordered (random) charge distribution is described by the Gaussian probability distribution function
\begin{equation}
   {\mathcal P}[\rho] = C\, \exp\bigg(\!\! - \frac{1}{2}  \int {\mathrm{d}} {\mathbf r}\,  g^{-1}({\mathbf r})\,
   								[\rho({\mathbf r}) - \rho_0({\mathbf r}) ]^2 \bigg),
\label{eq:pdf}								
\end{equation}
where 
$C$ is a normalization factor, $\rho_0({\mathbf r})$ is the mean  and $g({\mathbf r})$ the variance of the spatial distribution of random surface charges. It is obvious that the above probability distribution function entails an uncorrelated disorder, i.e., $ \langle \! \langle  \left(\rho({\mathbf r})-\rho_0({\mathbf r})\right)\left(\rho({\mathbf r}')-\rho_0({\mathbf r}')\right) \rangle   \!\rangle = g({\mathbf r})\delta({\mathbf r}-{\mathbf r}')$, where double-brackets denote the configurational (quenched) average $ \langle \! \langle  \cdots \rangle  \! \rangle = \int {\mathcal D}\rho \, {\mathcal P}[\rho]\, \big(\cdots\big)$.

In what follows, the general formalism is valid for an arbitrary shape of the charged boundaries, but for the later developments in this paper we shall delimit ourselves to the specific example of a planar slab as noted above. In this case, the mean charge distribution and its variance are given by
\begin{equation}
	\rho_0({\mathbf r}) = -\sigma e_0 \, \delta(z), \quad
	g({\mathbf r}) = g e_0^2 \, \delta(z),
\end{equation}
where $g\geq 0$, and, without loss of generality, we assume $\sigma\geq 0$ and, for the multivalent counterions,  $q\geq 0$.  The monovalent salt is assumed to be present on both sides of the slab while the multivalent counterions are restricted to the right half-space $z\geq 0$. This spatial constraint can be taken into account formally by introducing the indicator functions $\Omega_c({\mathbf r}) = \theta(z)$ for the multivalent counterions, and  $\Omega_+({\mathbf r}) =  \Omega_-({\mathbf r}) = \Omega_s({\mathbf r}) = \theta(z)+\theta(b-z)$ for the monovalent ions, where $ \theta(z)$ is the Heaviside's step function.  This constraint can be realized by enclosing the region on the left side of the slab in a membrane impermeable to multivalent ions; such a constraint will be relevant  in our analysis only in the situations where the slab thickness is small relative to the salt screening length and will otherwise have vanishing impact on the distribution of multivalent counterions next to the charged surface in the regime of parameters that will be of interest in this paper. For the most part, however, we shall focus on the case of semi-infinite slabs.

Another point to be noted here is that, for the sake of simplicity,  we model the monovalent and multivalent ions as point-like particles. Since the monovalent ions will be treated implicitly (Section \ref{sec:formal}),  the generalization of our formalism to include a finite size for the multivalent counterions (which can have a relatively large size as compared with the monovalent ions) is quite straightforward and we shall return to it in Section \ref{subsec:ion-size}, where we also analyze its consequences.

\section{Theoretical Background}
\label{sec:formal}

\subsection{Field action for dressed multivalent ions}

For a given realization of the fixed charge distribution $\rho({\mathbf r})$, the grand-canonical partition function of the above model can be written exactly in a functional-integral form as \cite{Edwards,Podgornik89,Podgornik89b,Netz01,Netz-orland}
\begin{equation}
  {\mathcal Z} = \rme^{-\frac{1}{2}\ln \,\det\, G} \int \!{\mathcal D}\phi \,\, \rme^{-\beta S[\phi]},
  \label{eq:Z}
\end{equation}
where $\beta = 1/(k_{\mathrm{B}}T)$ and $\phi({\mathbf r})$ is the fluctuating (electrostatic) potential and the effective ``field-action" reads
%
\begin{eqnarray}
\label{fieldaction}
  &&S[\phi] =  \frac{1}{2} \int \rmd {\mathbf r}\,  \rmd {\mathbf{r}'} \phi(\mathbf{r}) G^{-1}({\mathbf r}, {\mathbf r}') \phi(\mathbf{r}') +\nonumber\\
  &&\qquad\quad + \,\rmi \int \rmd {\mathbf r}\,\rho({\mathbf r})\phi({\mathbf r}) -  k_{\mathrm{B}}T \int \rmd {\mathbf r}\,{\mathcal V}(\phi(\mathbf{r})),
\end{eqnarray}
where $G^{-1}({\mathbf r}, {\mathbf r}')$ is the operator inverse of the Coulomb interaction (or the bare Green's function), $G({\mathbf r}, {\mathbf r}')$, in the presence of dielectric boundaries satisfying the equation $-\epsilon_0 \nabla\cdot \epsilon({\mathbf r}) \nabla G({\mathbf r}, {\mathbf r}')  = \delta({\mathbf r} -{\mathbf r}')$. The ``field self-interaction" term is given by  \cite{SCdressed1,perspective}
 \begin{equation}
 {\mathcal V}(\phi)\! = \! \lambda_c\Omega_c({\mathbf r})  \rme^{-\rmi\beta q e_0  \phi} +\Omega_s({\mathbf r})\left(\lambda_+ \rme^{-\rmi\beta e_0  \phi} + \lambda_-  \rme^{\rmi  \beta e_0  \phi}\right),
  \label{eq:V_s}
\end{equation}
where $\lambda_c$ and $\lambda_{\pm}$ are the ionic fugacities of multivalent counterions ($c$) and monovalent anions ($-$) and cations ($+$), respectively.
Note that, since the multivalent counterions result from a $q$:1 salt of  bulk concentration $c_0$ mixed with a 1:1 salt of bulk concentration $n_0$, we have $\lambda_c = c_0$,  $\lambda_+=n_0$ and $\lambda_- = n_0+qc_0$.

The above field action obviously represents  a highly asymmetric Coulomb fluid especially when  counterions are multivalent, $q > 1$. These counterions   couple strongly to the fixed (mean) surface charge, whereas the monovalent salt species couple weakly.  This leads to a complex situation where different components of the Coulomb fluid couple differently to the same surface charges, thus making the analytical progress and, in particular, obtaining exact solutions \cite{exact1,exact2}, very difficult. 

Nevertheless, progress is possible and systems of this type can be treated using a combined weak-strong coupling approximation, which has been discussed in a series of recent works \cite{SCdressed1,SCdressed2,SCdressed3,perspective}. It was shown, by employing both analytical approaches as well as implicit- and explicit-ion simulations that, in a wide range of realistic system parameters, the monovalent ions can be treated safely within the Debye-H\"uckel (DH) framework, while the multivalent ions can be handled by means of a standard virial expansion scheme \cite{Netz01,AndrePRL,AndreEPJE,hoda_review,perspective,Naji_PhysicaA,asim} of the strong coupling approximation. The DH-type terms in this context follow by expanding the last two terms in Eq. (\ref{eq:V_s}) up to the second order in $\phi({\mathbf r})$ (which can be justified on a systematic basis  in highly asymmetric systems with $q \gg 1$  \cite{SCdressed1}), i.e., 
\begin{equation}
{\mathcal V}(\phi({\mathbf r})) \simeq \lambda_c  \Omega_c({\mathbf r})\, \rme^{-\rmi \beta q e_0  \phi} -  n_b \Omega_s({\mathbf r})  (\beta e_0  \phi)^2/2 + {\cal O}(\phi^3),
\end{equation}
where $n_b$ is the bulk concentration due to all monovalent ions $n_b= \lambda_+ + \lambda_- = 2n_0+qc_0$.

On the analytical level, the above procedure allows one to trace the partition function over the degrees of freedom associated with monovalent ions and thus one remains with only ``dressed" multivalent ions and surface charges that then interact through a DH-type interaction kernel (or the screened Green's function) defined via
\begin{equation}
-\epsilon_0 \nabla\cdot \epsilon({\mathbf r}) \nabla G({\mathbf r}, {\mathbf r}')  + \epsilon_0  \epsilon({\mathbf r})\kappa^2({\mathbf r})G({\mathbf r}, {\mathbf r}')= \delta({\mathbf r} -{\mathbf r}'),
\label{eq:G_DH}
\end{equation}
where the Debye (or salt) screening parameter $\kappa({\mathbf r})$ is non-zero only outside the dielectric slab and is given by  $\kappa^2 = 4\pi \ell_{\mathrm{B}} n_b$ with $\ell_{\mathrm{B}}= e_0^2/(4\pi \epsilon_0 \epsilon_m k_{\mathrm{B}}T)$ being the Bjerrum length. 

This type of methodology leads to the so-called {\em dressed multivalent-ion theory} \cite{SCdressed1,SCdressed2,SCdressed3,perspective}, which is  a direct generalization of the standard counterion-only SC theory \cite{Netz01,AndrePRL,AndreEPJE,hoda_review,perspective,Naji_PhysicaA,asim}. In fact, the dressed multivalent-ion theory has a hybrid character in that it reproduces both the counterion-only SC theory and the DH theory as two asymptotic limits at small and large salt screening parameters, respectively. 

In what follows, we shall use this framework to study the effects of surface charge disorder on the distribution of multivalent counterions.

\subsection{Counterion density profile}
\label{subsec:ci_dens}

The density profile of multivalent counterions follows standardly from the general formalism defined by Eqs. (\ref{eq:Z}) and (\ref{fieldaction}) as \cite{Netz01,Netz-orland,AndreEPJE,hoda_review,perspective,Naji_PhysicaA,AndrePRL,asim}
\begin{equation}
  c({\mathbf r}; [\rho])  =   \lambda_c \Omega_c({\mathbf r})\langle \rme^{-\rmi\beta q e_0  \phi}\rangle,
 \end{equation}
where $\langle\cdots \rangle$ denotes the thermal (ensemble) average over the fluctuating field $\phi({\mathbf r})$. For a given (quenched) realization of $\rho({\mathbf r})$, this average can be calculated analytically in both limits of weak and strong coupling in the counterion-only case \cite{Netz01,Netz-orland} and also in the more general context of dressed multivalent ions in asymmetric Coulomb fluids \cite{SCdressed1,SCdressed2,SCdressed3}. As noted before,  the effects due to multivalent ions can be investigated
by virial expanding the partition function in terms of their fugacity and  keeping only the leading order contributions (see Refs. \cite{Netz01,hoda_review,perspective,Naji_PhysicaA,SCdressed1,SCdressed2,SCdressed3,AndrePRL,AndreEPJE,asim} for further details). This procedure leads to a single-particle form for the density profile of multivalent counterions as
\begin{equation}
	\label{eq:sc_density}
         c({\mathbf r}; [\rho])  =   \lambda_c \Omega_c({\mathbf r}) \, \rme^{-\beta u({\mathbf r}; [\rho])},
\end{equation}
where $u({\mathbf r}; [\rho])$ is the single-particle interaction energy
\begin{equation}
  u({\mathbf r}; [\rho]) = qe_0 \int {\mathrm{d}}{\mathbf r}' \,G({\mathbf r}, {\mathbf r}')\rho({\mathbf r}') + \frac{q^2e_0^2}{2} G_{\mathrm{im}}({\mathbf r}, {\mathbf r}).
  \label{eq:u_rho}
\end{equation}
Here, $G_{\mathrm{im}}({\mathbf r}, {\mathbf r})$ is the generalized Born energy contribution that stems purely from the dielectric and/or salt polarization effects (or the so-called  ``image charges"). In other words, $G_{\mathrm{im}}({\mathbf r}, {\mathbf r})= G({\mathbf r}, {\mathbf r}) - G_0({\mathbf r}, {\mathbf r})$, where  $G_0({\mathbf r}, {\mathbf r})$ is the formation (self-)energy of individual counterions in a homogeneous background, which is obtained from the free-space screened Green's function defined via $-\epsilon_0 \epsilon_m(\nabla^2 - \kappa^2) G_0({\mathbf r}, {\mathbf r}')  = \delta({\mathbf r} -{\mathbf r}')$.

In the present model, the fixed surface charge distribution, $\rho({\mathbf r})$, has a disorder component and thus, in order to obtain the measurable counterion density profile, one must average Eq. (\ref{eq:sc_density}) over different realizations of this disorder field using the Gaussian weight  (\ref{eq:pdf}). This can be done straightforwardly by computing
\begin{equation}
         c({\mathbf r})  =  \langle \! \langle c({\mathbf r}; [\rho]) \rangle\! \rangle =  \lambda_c \Omega_c({\mathbf r}) \langle \! \langle \rme^{-\beta u({\mathbf r}; [\rho])} \rangle\! \rangle,
\end{equation}
which  then gives
\begin{equation}
	\label{eq:sc_density_av}
         c({\mathbf r})  =    \lambda_c \Omega_c({\mathbf r}) \, \rme^{-\beta u({\mathbf r})},
\end{equation}
where the {\em effective} single-particle interaction energy now reads 
\begin{equation}
u({\mathbf r})= u_0({\mathbf r})+ u_{\mathrm{im}}({\mathbf r}) + u_{\mathrm{dis}}({\mathbf r}),
\label{eq:u}
\end{equation}
with
\begin{eqnarray}
  &&u_0({\mathbf r}) = qe_0 \int {\mathrm{d}}{\mathbf r}'\, G({\mathbf r}, {\mathbf r}')\rho_0({\mathbf r}'), \\
   &&u_{\mathrm{im}}({\mathbf r}) = \frac{q^2e_0^2}{2} G_{\mathrm{im}}({\mathbf r}, {\mathbf r}), \\
   && u_{\mathrm{dis}}({\mathbf r}) = - \beta  \frac{q^2e_0^2}{2} \int {\mathrm{d}}{\mathbf r}' g(\Av r') [G(\Av r,\Av r')]^2.
   \label{eq:u2}
\end{eqnarray}
The three terms, respectively, originate from the contribution of the interaction of multivalent ions with the mean surface charge density (first term), the contribution of self-interactions (interactions with image charges, second term) and the contribution of the surface charge disorder (third term). The latter can be viewed as an effective surface-counterion interaction which is induced by the quenched  randomness  in the surface charge. It is proportional to the disorder variance but also shows an explicit {\em temperature dependence}. Another interesting point is that the disorder interaction term exhibits a quadratic dependence on the Green's function and likewise also on the multivalent ion charge valency $q$. These features can be understood by noting that the disorder term in fact represents the {\em sample-to-sample fluctuations (or variance)} of the single-particle interaction energy, $u({\mathbf r}; [\rho])$,  Eq. (\ref{eq:u_rho}), which is a linear functional of the Gaussian field $\rho({\mathbf r})$, and hence one can show that 
\begin{equation}
u({\mathbf r})= \langle \! \langle  u({\mathbf r}; [\rho])\rangle\! \rangle - \beta \langle \! \langle u^2({\mathbf r}; [\rho])\rangle\! \rangle_c/2, 
\label{eq:u_v2}
\end{equation}
where $\langle \! \langle u^2({\mathbf r}; [\rho])\rangle\! \rangle_c\equiv   \langle \! \langle u^2({\mathbf r}; [\rho])\rangle\! \rangle - \langle \! \langle  u({\mathbf r}; [\rho])\rangle\! \rangle^2$, and 
\begin{eqnarray}
  &&\langle \! \langle  u({\mathbf r}; [\rho])\rangle\! \rangle = u_0({\mathbf r})+ u_{\mathrm{im}}({\mathbf r}), \\
   &&\langle \! \langle u^2({\mathbf r}; [\rho])\rangle\! \rangle_c=  -2k_{\mathrm{B}}T\,u_{\mathrm{dis}}({\mathbf r}).
\end{eqnarray}

\subsection{Rescaled representation}

In order to proceed, we introduce the dimensionless (rescaled) quantities 
\begin{equation}
\tilde{\mathbf r}={\mathbf r}/\mu, \quad \tilde\kappa=\kappa\mu, \quad \tilde b=b/\mu, \quad \Xi = q^2\ell_{\mathrm{B}}/\mu,
\end{equation}
where
\begin{equation}
\mu = 1/(2\pi q \ell_{\mathrm{B}}  \sigma)
\label{eq:mu}
\end{equation}
is the Gouy-Chapman length and $\Xi =  2\pi q^3\ell_{\mathrm{B}}^2\sigma$ is the electrostatic coupling parameter associated with the mean surface charge \cite{Netz01,AndreEPJE,hoda_review,perspective,Naji_PhysicaA,AndrePRL,asim}. Analogously, one can define the dimensionless {\em disorder coupling (or disorder strength) parameter} 
\begin{equation}
  \chi = 2\pi q^2\ell_{\mathrm{B}}^2 g,
\end{equation}
which is proportional to the disorder variance; its dependence on the counterion valency, $q$, is different from that of the mean electrostatic coupling parameter as first noted in Ref. \cite{ali-rudi}.

\section{Results}
\label{sec:results}

\subsection{Counterion-only case}
\label{subsec:ci-only}

Let us first consider the special case of multivalent counterions next to a charged surface in the absence of any salt screening ($\kappa=0$) and dielectric image charge effects (i.e., in  a dielectrically homogeneous system  with $\epsilon_p=\epsilon_m$).  This case has been considered in a previous work on the effective interaction between two randomly charged surfaces \cite{ali-rudi}, which however did not investigate the distribution of counterions. 

In this case, the number of counterions, $N$, is fixed by the mean charge on the surface through the electroneutrality condition  $Nq=S\sigma$, where $S$ is the surface area. The fugacity of counterions is thus given by \cite{ali-rudi}
 \begin{equation}
\lambda_c =\frac{N}{ \int{\mathrm{d}}{\mathbf r}\, \Omega_c (\Av r) \, \rme^{-\beta u(\Av r)}}.
\label{eq:lambda_c}
\end{equation}
The effective single-particle interaction is obtained by using Eqs. (\ref{eq:u})-(\ref{eq:u2}) and noting that, in this case, $G({\mathbf r}, {\mathbf r}') = 1/(4\pi \epsilon_0\epsilon_m |{\mathbf r} - {\mathbf r}'|)$. Hence,  in rescaled units, we have $\beta u_{\mathrm{im}}({\mathbf r}) = 0$, and
\begin{eqnarray}
  &&\beta u_0(\tilde z) = \tilde z, \\
   &&\beta u_{\mathrm{dis}}(\tilde z) = \frac{\chi}{2}\ln\,\tilde z
   \label{eq:u2_ci_only}
\end{eqnarray}
 (up to irrelevant additive constants). Therefore, 
\begin{equation}
\beta u(\tilde z)= \tilde z +\frac{\chi}{2}\ln\,\tilde z.
\label{eq:u_z}
 \end{equation}
The rescaled density profile of counterions is then obtained using Eq. (\ref{eq:sc_density_av}) as
 \begin{equation}
\tilde c(\tilde z)\equiv\frac{c(\tilde z)}{2\pi \ell_{\mathrm{B}} \sigma^2}=\frac{\tilde z^{-\frac{\chi}{2}}\;\rme^{-\tilde z}}{\Gamma (1-\frac{\chi}{2})}, \qquad \tilde z\geq 0, 
\label{eq:dens_0}
\end{equation}
where $\Gamma(\cdot)$ is the Gamma function. 
This expression shows a standard SC exponential decay of the multivalent counterion density, which dominates at large separations from the surface and is a well-established result within the SC context \cite{Netz01,AndreEPJE,hoda_review,perspective,Naji_PhysicaA,AndrePRL,asim,Shklovs02}. But it also exhibits an algebraic  dependence on $\tilde z$, which dominates at small separations from the surface and thus shows that, in the presence of surface charge disorder, the counterion density {\em diverges} in the immediate vicinity of the surface. The presence of disorder thus clearly violates the contact-value theorem which was derived for uniformly charged surfaces \cite{contact_value,contact_value2,contact_value1,contact_value3}; this theorem entails a contact value of  $\tilde c(0) =1$ in the whole range of coupling parameters. Note that, nevertheless, the electroneutrality is exactly satisfied as $\int_0^{\infty }\rmd\tilde z\,\tilde c(\tilde z)  = 1$.

\begin{figure}[t!]
\begin{center}
	\includegraphics[width=8cm]{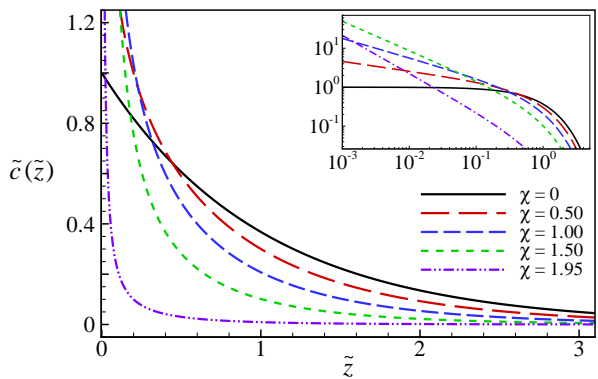}
\caption{(Color online) Rescaled density profile of multivalent counterions next to a randomly charged surface in the absence of salt and image charge effects for different values of the disorder coupling parameter as shown on the graph. Inset shows the diverging behavior of the density profile at small distances from the surface in the log-log scale. 
}
\label{f:densityp_fin}
\end{center}
\end{figure}

The behavior of the density profile, Eq. (\ref{eq:dens_0}),  is shown in Fig. \ref{f:densityp_fin} for a few different values of the disorder coupling parameter. As seen, due to the singular behavior at the surface, the presence of charge disorder  enhances (suppresses)  the density of counterions at small (large) separations, in general agreement with the previous findings in the case of non-disordered but heterogeneously charged surfaces such as surfaces carrying discrete charge patterns (see, e.g., \cite{Andre02,Lukatsky,pincus,Nikoofard} and references therein).

\begin{figure}[t!]
\begin{center}
	\includegraphics[width=8cm]{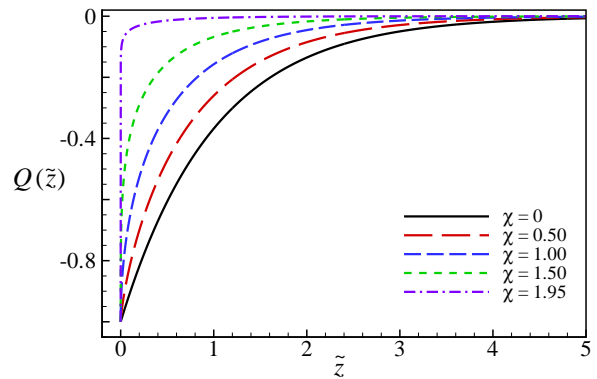}
\caption{(Color online) Cumulative charge next to a randomly charged surface in the absence of salt and image charge effects for different values of the disorder coupling parameter as shown on the graph.
}
\label{f:Q_densityp_fin}
\end{center}
\end{figure}

The above results can be illuminated further by analyzing the averaged cumulative charge defined as
\begin{equation}
Q(z)=\frac {1}{\sigma}\! \int_0^z\!\!\rmd z'\left[ e_0q c(z')+\rho_0(z')\right],
\end{equation}
or, in rescaled units, 
\begin{equation}
  Q({\tilde z}) = -1+\int_0^{\tilde z} \rmd \tilde z' \, \tilde c({\tilde z'}) = - \frac{\Gamma(1-\frac{\chi}{2}, {\tilde z})}{\Gamma(1-\frac{\chi}{2})}, 
  \label{eq:Q0}
\end{equation}
where $\Gamma(\cdot,\cdot)$ is the incomplete Gamma function. 
Note that $Q(z)$  is normalized such that $Q(0) =-1$, which  represents  the mean surface charge density at $\tilde z = 0$, and  $Q(z)\rightarrow 0$ for $z\rightarrow\infty$, which reflects the global electroneutrality of the system.  As seen in Fig. \ref{f:Q_densityp_fin}, nearly all of the counterions  become strongly localized in the vicinity of the surface (i.e., $Q({\tilde z})\simeq 0$ for finite $\tilde z>0$) as  $\chi$ is increased. In fact, the counterion density profile tends to zero at any finite separation from the surface $\tilde c(\tilde z)\rightarrow 0$ when $\chi$ tends (from below) to the threshold value
\begin{equation}
\chi_* = 2,
\end{equation}
because the Gamma function in the denominator of the normalization factor in Eqs. (\ref{eq:dens_0}) and (\ref{eq:Q0}) goes to infinity. 

It should be emphasized that the strong accumulation of multivalent counterions in the immediate vicinity of the surface does not give rise to a renormalized mean surface charge density (see also Appendix C in Ref. \cite{partial}) and that the singular behavior of counterions at the surface in the present context should be distinguished from the surface adsorption or counterion condensation phenomena \cite{perspective}. 

\subsection{Disorder-induced anti-fragility}

The foregoing results clearly show that the excess accumulation of counterions near the surface is driven by the disorder-induced, single-particle interaction energy (\ref{eq:u2_ci_only}), which is attractive and depends logarithmically on the distance from the surface and thus generates the singular behavior of the counterion density profile at the surface. This suggests that the presence of quenched surface charge disorder drives the system towards a state of lower thermal `disorder'. This point can be established systematically by calculating the difference in the entropy of counterions in the presence and in the absence of disorder, $\Delta S(\chi)=S(\chi)-S(0)$. We find
\begin{equation}
  \frac{\Delta S(\chi)}{Nk_{\mathrm{B}} }= \chi \psi\left(1-\frac{\chi}{2}\right)+\ln\, \Gamma\left(1-\frac{\chi}{2}\right), 
\end{equation}
where $\psi(\cdot)$ is the digamma function. It thus follows that $\Delta S(\chi)\leq 0$. The entropy reduction is larger for larger disorder strength, $\chi$, and diverges as $\chi\rightarrow 2^-$. 

In other words, the reduction  in the {\em translational entropy} of multivalent counterions in the solution is driven by introducing a finite degree of {\em configurational entropy} due to the presence of quenched randomness in the surface charge distribution. Formally, this latter type of entropy is generated by the non-thermal (quenched) average taken over different realizations of the  surface charge disorder. This subtle interplay between the different kinds of entropy is therefore essential in generating the  singular behavior of counterions near the disordered surface. 
It thus seems appropriate to refer to this type of behavior of the multivalent counterions as {\em anti-fragile} \cite{taleb}, since introducing an external (quenched) disorder source effectively diminishes the intrinsic thermal disorder in the system and drives it towards a more `ordered' state. 

Another interesting point to be noted here is that the internal energy of the system also decreases due to the presence of disorder but in such a way that leads to a decrease in the free energy of the system as can be seen from the free energy difference, $\beta \Delta {\mathcal F}(\chi)/N =-\ln\, \Gamma\left(1-{\chi}/{2}\right)\leq 0$. Therefore, the system also attains a thermodynamically more stable state, which is again a direct consequence of the singular behavior of counterions near the disordered surface. By contrast, one can show that in the case of counterions next to a uniformly charged surface, which exhibits a regular potential, the attraction of counterions towards the surface leads to a larger free energy as compared with the ideal case where the system is uncharged.

\subsection{Salt image effects}
\label{subsec:salt}

We now turn to  the effects due to a monovalent salt bath by assuming that in addition to the multivalent counterions (of bulk concentration $c_0$), we also have a finite amount of  monovalent  salt  in the system, giving a total bulk monovalent ion concentration of $n_b=2n_0+qc_0$ (Sections \ref{sec:model} and \ref{sec:formal}). We take a semi-infinite slab  ($ b =\infty$) impermeable to all ions and  assume that the system is again  dielectrically homogeneous, i.e. $\epsilon_p=\epsilon_m$. This helps to disentangle the polarization effects due to  the inhomogeneous distribution of salt ions (``salt image effects")  from those resulting from the inhomogeneous distribution of the dielectric constant (``dielectric image effects").  

For a semi-infinite slab, the Fourier-Bessel transform of the Green's function can be obtained by standard methods as
\begin{equation}
   \hat G(Q;z,z')= \frac{1}{2\epsilon _0\epsilon _m\gamma }\left[\rme^{-\gamma |z-z'|}+\Delta_s  \,\rme^{-\gamma (z+z')}\right],
   \label{eq:inf_green}
\end{equation}
where
\begin{equation}
 \Delta_s = \frac{\gamma- Q }{\gamma + Q}, \qquad  \gamma^2=Q^2+\kappa^2.
\end{equation}
Note that because of the translational and rotational  symmetry  with respect to the transverse (in-plane) coordinates $\Avg\rho = (x, y)$ and $\Avg\rho' = (x', y')$, the Green's function depends on these coordinates  only through   $|\Avg\rho-\Avg\rho'|$, i.e., $G(\Av r,\Av r') = G(|\Avg\rho- \Avg\rho'|;z, z')$,  and  thus its Fourier-Bessel transform  is defined via  $G(\Av r-\Av r')=\int_0^\infty \frac{Q {\mathrm{d}}Q}{2\pi }\,\hat G(Q;z,z')\,J_0(Q \vert \Avg\rho -\Avg\rho'\vert)$. 

\begin{figure}[t!]
\begin{center}
\includegraphics[width=8.cm]{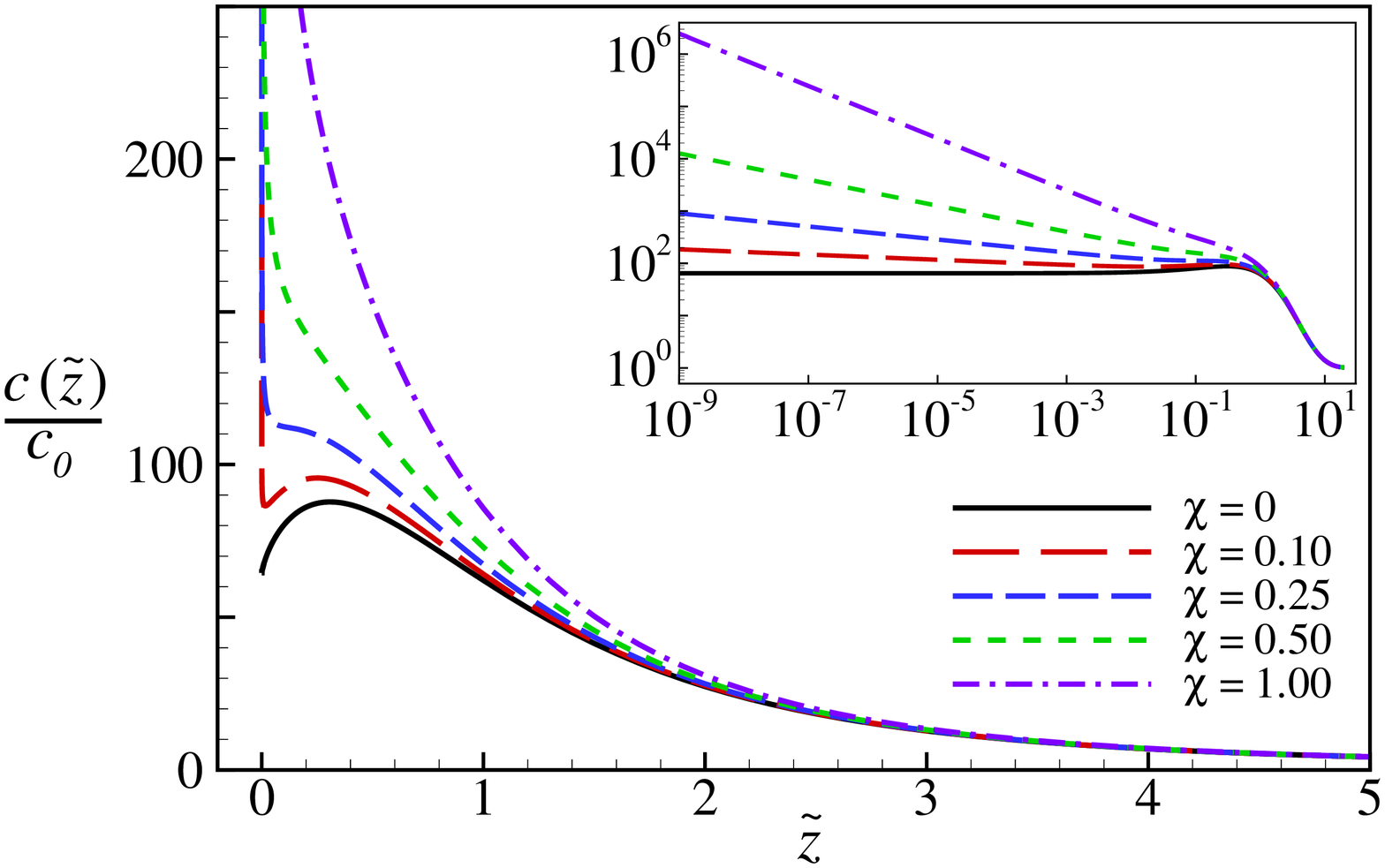}
\caption{(Color online) Rescaled density profile of multivalent counterions next to the randomly charged surface of a semi-infinite slab for $\Xi=50$,  $\tilde \kappa=0.3$ and in the absence of dielectric image charge effects ($\epsilon_p=\epsilon_m$). Different curves correspond to  different values of the disorder coupling parameter as shown on the graph.  Inset shows the diverging behavior of the density profile at small distances from the surface in the log-log scale. 
}
\label{f:contact_inf}
\end{center}
\end{figure}

The contributions to the effective single-particle interaction energy, $u=u_0+u_{\mathrm{im}}+u_{\mathrm{dis}}$, Eqs. (\ref{eq:u})-(\ref{eq:u2}), are now obtained, in rescaled units, as
\begin{eqnarray}
\label{eq:inf_u0}
&& \beta u_0(\tilde z)=-\frac{2}{\tilde \kappa}\rme^{-\tilde \kappa\tilde z},\\
&& \beta u_{\mathrm{im}}(\tilde z) =\frac{\Xi}{2}\int_0^\infty \tilde Q {\mathrm{d}}\tilde Q\, \frac{\Delta_s}{\tilde\gamma}\rme^{-2\tilde\gamma\tilde z}, \\
&& \beta u_{\mathrm{dis}}(\tilde z)=-\frac{\chi}{2}\int_0^\infty \tilde Q {\mathrm{d}}\tilde Q \,\frac{(1+\Delta_s)^2}{\tilde\gamma ^2}\rme^{-2\tilde\gamma\tilde z},
\label{eq:inf_u2}
\end{eqnarray}
where $\tilde \gamma = \gamma \mu$. These equations  can be used along with Eq. (\ref{eq:sc_density_av}) in order to compute the density profile of counterions. 

As seen in Fig. \ref{f:contact_inf}, the rescaled density profile of counterions shows a clear depletion effect in the absence of disorder (note that here we have rescaled the density of counterions with  their bulk concentration). This behavior is caused by the salt image effects that are produced by the second term in Eq. (\ref{eq:inf_green}). The depletion effect becomes weaker when the surface is randomly charged as the counterions are again attracted more strongly to the surface in the presence of charge disorder. The interplay between salt image depletion and disorder attraction leads to a non-monotonic behavior in the counterion density profile as the disorder effects dominate at small separations while the salt image effects dominate at intermediate separations. At large distances from the surface ($\tilde z\gtrsim \tilde \kappa^{-1}$), the behavior of the density profile is dominated by the salt screening effects and we get $c(\tilde z)/c_0\rightarrow 1$ for $\tilde z\rightarrow \infty$. The accumulation of counterions in the vicinity of the surface is  suppressed when the  bulk salt concentration is increased  or when a finite dielectric discontinuity is introduced at the surface (see Section \ref{subsec:delta}). 

Note that the counterion density profiles  show a singular bahavior  at the surface even in the presence of additional salt. This behavior is in fact present at {\em any} finite values of $\chi$ (as may be discerned more clearly from the log-log plot in the inset of Fig. \ref{f:contact_inf}) and coincides with the same algebraic divergence $\sim \tilde z^{-\chi/2}$ on approach to the surface as we found in the counterion-only case in the previous Sections. This is intuitively expected because the salt bath effects diminish at separations much smaller than the screening length $\tilde \kappa^{-1}$. 

\begin{figure}[t!]
\begin{center}
\includegraphics[width=8cm]{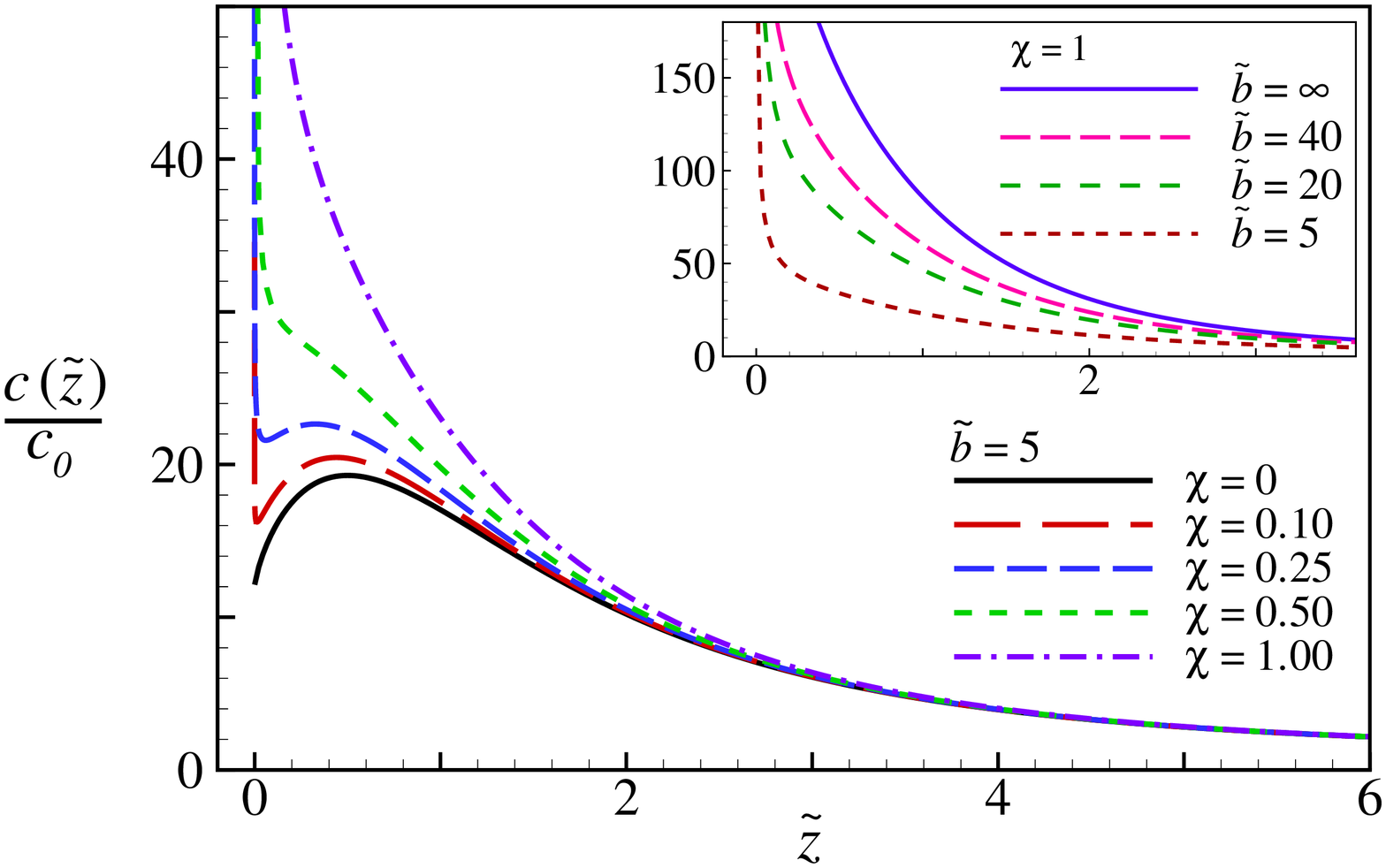}
\caption{(Color online) Same as Fig. \ref{f:contact_inf} but for fixed rescaled slab thickness $\tilde b = 5$ and different values of the disorder coupling parameter  as shown on the graph (main set) and also for fixed disorder coupling parameter $\chi=1$ and different values of the rescaled slab thickness (inset).}
\label{f:contact_inf_b}
\end{center}
\end{figure}

The effects of a finite slab thickness, $b$, can be examined by using the appropriate form of the Green's function in this case that can be obtained by means of standard methods as
\begin{eqnarray}
&&\hat G(Q;z,z')= \frac{1}{2 \epsilon _0\epsilon _m\gamma }\times\nonumber\\
  &&\quad\quad \times \left[\rme^{-\gamma |z-z'|}+\frac{\Delta_s \left(1-\rme^{-2Q b}\right)}{1-\Delta_s^2\,\rme^{-2Q b}}\rme^{-\gamma (z+z')}\right].
\end{eqnarray}
Hence, the three different terms in the effective single-particle interaction energy follow in rescaled units as
\begin{eqnarray}
\label{eq:b_u0}
&& \beta u_0(\tilde z)=- \frac{2(1+   \tilde\kappa\tilde b )}{\tilde\kappa(2  +   \tilde\kappa\tilde b)}\rme^{-\tilde\kappa\tilde z},  \\
&& \beta u_{\mathrm{im}}(\tilde z) =\frac{\Xi}{2}\!\int_0^\infty\!\! \tilde Q {\mathrm{d}}\tilde Q\, \frac{\Delta_s  \left(1-\rme^{-2\tilde Q \tilde b}\right)}{\tilde\gamma\left(1-\Delta_s^2\,\rme^{-2\tilde Q \tilde b}\right)}\rme^{-2\tilde\gamma\tilde z}, 
\label{eq:b_u1}
\\
&& \beta u_{\mathrm{dis}}(\tilde z)=-\frac{\chi}{2}\!\int_0^\infty\!\! \tilde Q {\mathrm{d}}\tilde Q \,\frac{(1+\Delta_s)^2\left(1-\Delta_s \, \rme^{-2\tilde Q\tilde b}\right)^2}{\tilde\gamma ^2\left(1-\Delta_s ^2\, \rme^{-2\tilde Q \tilde b}\right)^2}\rme^{-2\tilde\gamma\tilde z}. \nonumber\\
\label{eq:b_u2}
\end{eqnarray}
The finiteness of the slab thickness is expected to be relevant mainly in the regime where the thickness is comparable with or smaller than the screening length, i.e. $\tilde \kappa  \tilde b\lesssim1$. As it can be seen directly from Eqs. (\ref{eq:b_u0})-(\ref{eq:b_u2}), both the attraction experienced by multivalent counterions due to the mean surface charge and its disorder variance and the repulsion due to the salt image effects become stronger as the slab becomes thicker or, in other words,  as the system becomes more strongly inhomogeneous in terms of the salt distribution. The overall effect is such that the counterion density close to the charged surface becomes smaller for smaller $\tilde b$ as shown in Fig. \ref{f:contact_inf_b}, inset (compare also Figs. \ref{f:contact_inf} and \ref{f:contact_inf_b}). This behavior can be understood also by noting that for  thinner slabs the salt ions on the left side of the slab also contribute to the screening effects and, hence, further suppress the multivalent counterion density on the right side (see Fig. \ref{f:schematic}). This however leaves the effects resulting from the surface charge disorder qualitatively unchanged, especially at small distances from the surface, where the singular behavior persists.

It is to be noted that the expressions for the  effective single-particle interaction energy, Eqs.  (\ref{eq:b_u0})-(\ref{eq:b_u2}), can correctly reproduce the counterion-only result, Eq. (\ref{eq:u_z}), when the limit $\tilde\kappa\rightarrow 0$, which also gives $\Delta_s\rightarrow 0$, is taken (in this case, the thickness $\tilde b$ will be irrelevant). The counterion-only limit cannot be recovered if we start with the infinite-thickness expressions  (\ref{eq:inf_u0})-(\ref{eq:inf_u2}) (where the thickness of the slab is strictly set equal  to infinity)  and then take the limit of zero salt. The difference would be in a factor 2 in the expression for $u_0$,  indicating that the two limits  $\tilde \kappa\rightarrow 0$ and  $\tilde b\rightarrow \infty$ do not commute. Alternatively,  one can  recover Eq. (\ref{eq:u_z}) from Eqs.  (\ref{eq:b_u0})-(\ref{eq:b_u2})  by first taking the limit of a thin slab $\tilde b\rightarrow 0$ and then $\tilde\kappa\rightarrow 0$.

\begin{figure}[t!]\begin{center}
	\begin{minipage}[b]{0.44\textwidth}\begin{center}
		\includegraphics[width=\textwidth]{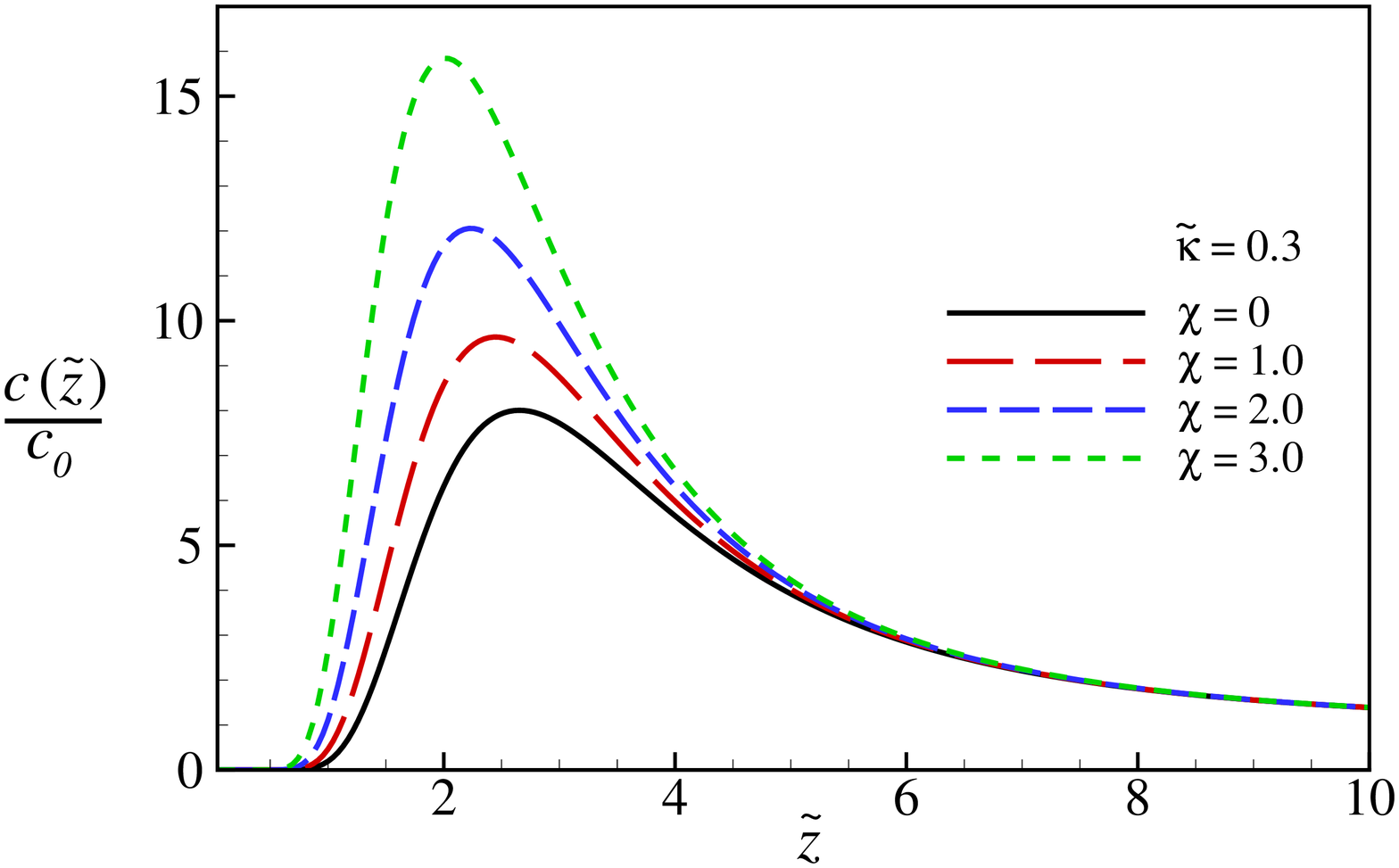} (a)
	\end{center}\end{minipage}\vskip.3cm
	\begin{minipage}[b]{0.44\textwidth}\begin{center}
		\includegraphics[width=\textwidth]{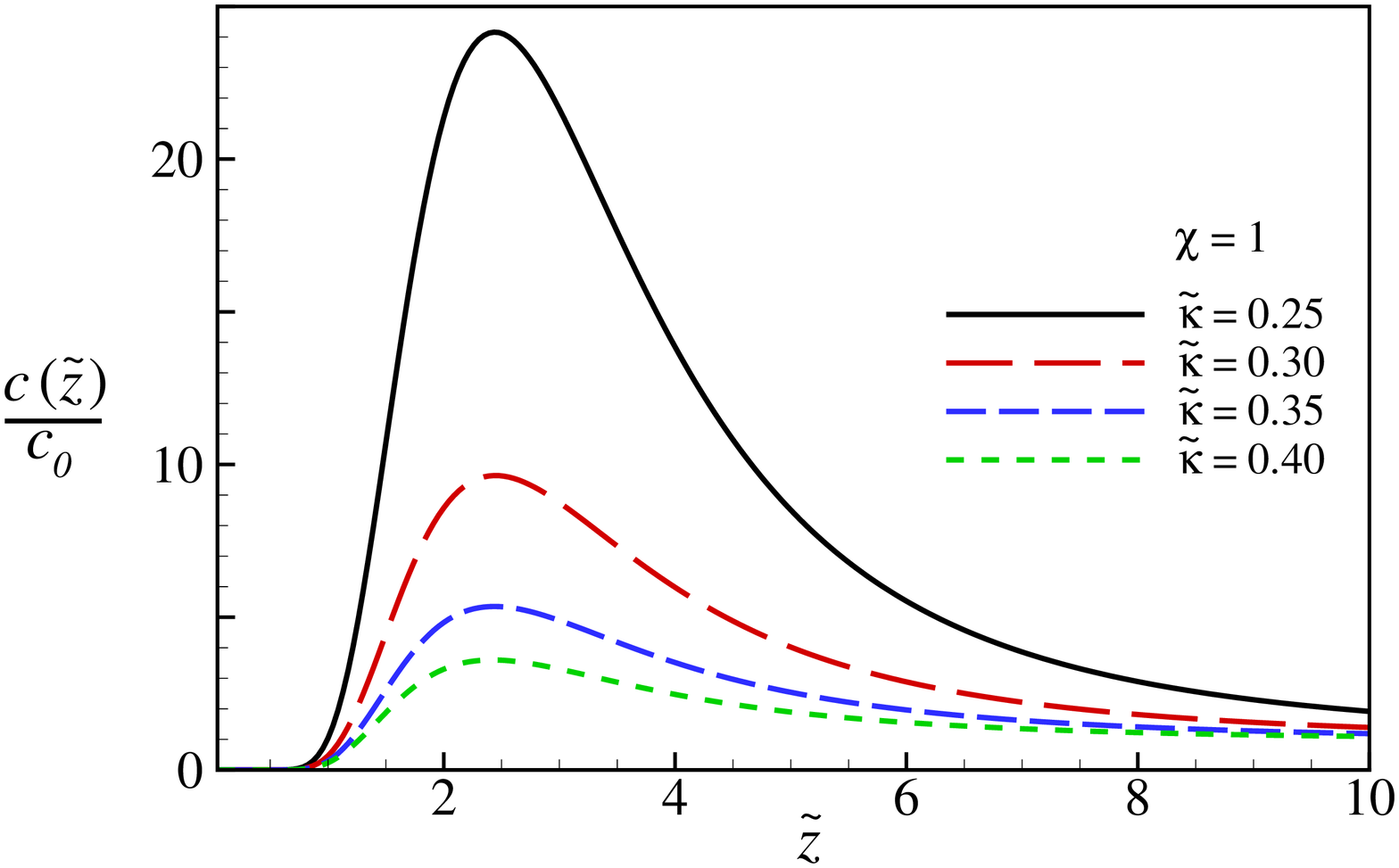} (b)
	\end{center}\end{minipage} 	
\caption{(Color online) (a) Rescaled density profile of multivalent counterions next to the randomly charged surface of a semi-infinite slab for $\Xi=50$,  $\tilde \kappa=0.3$, $\Delta=0.95$ and different values of the disorder coupling parameter as shown on the graph. (b) Same as panel (a) but for $\chi=1$ and different values of the rescaled  salt screening parameter. }
\label{fig:density_inf_delta}
\end{center}
\end{figure}

\subsection{Dielectric image effects}
\label{subsec:delta}

So far we focused on the cases with no dielectric discontinuity at the boundaries of the slab. The case of a dielectrically inhomogeneous system with $\epsilon_p \neq\epsilon_m$ can be studied by simply replacing the definition of $\Delta_s$ in the expressions in the previous section (e.g., Eqs. (\ref{eq:inf_u0})-(\ref{eq:inf_u2}))  with a more general one, i.e.
\begin{equation}
 \Delta_s = \frac{\epsilon_m\gamma- \epsilon_p Q }{\epsilon_m\gamma + \epsilon_p Q}.
 \label{eq:delta_s_general}
\end{equation}
In the absence of a salt bath ($\tilde \kappa=0$), the dielectric image charges lead to very strong repulsions from the surface when $\epsilon_p <\epsilon_m$ (as is often the case for aqueous solvents and macromolecular surfaces). This effect enters through the image interaction term $\beta u_{\mathrm{im}} = \Xi\Delta/4\tilde z$, where the dielectric discontinuity parameter is defined as
\begin{equation}
 \Delta = \frac{\epsilon_m- \epsilon_p  }{\epsilon_m + \epsilon_p}.
\end{equation}
The image interaction term diverges at the surface and thus implies a vanishing contact density $c(\tilde z)\rightarrow 0$  for $\tilde z\rightarrow 0$, a behavior that is very distinct from that generated merely by salt images (Figs. \ref{f:contact_inf} and \ref{f:contact_inf_b}), as the latter  cannot be described generally in terms of ``point-like image charges" and generate much weaker repulsive forces on multivalent ions than the dielectric images.

In the most general case with both salt and dielectric image effects, the density profile of counterions can be calculated via Eqs. (\ref{eq:sc_density_av}) and (\ref{eq:inf_u0})-(\ref{eq:inf_u2}) with the definition in Eq. (\ref{eq:delta_s_general}). The results are shown in Fig. \ref{fig:density_inf_delta} for $\tilde \kappa=0.3$  and $\Delta=0.95$, corresponding to the dielectric discontinuity at the water/hydrocarbon interface (with $\epsilon_p=2$ and $\epsilon_m=80$). Clearly, both  added salt  and charge disorder effects become irrelevant in the small-distance regime, where the dielectric-image repulsions dominate and generate a wide depletion zone near the surface. The counterion density thus again vanishes at the surface and tends to the bulk value at large separations. Although the qualitative form of the density profile remains the same in the absence and in the presence of charge disorder (which is in contrast with what we found in Section \ref{subsec:salt}), the peak of the density profile becomes more pronounced and shifts to smaller values of the distance from the surface as the disorder strength is increased, see Fig. \ref{fig:density_inf_delta}a. A similar effect is seen in Fig. \ref{fig:density_inf_delta}b, where the salt screening parameter is decreased; in this case the location of the peak of the density profile remains nearly unchanged while its  height increases by almost an order of magnitude when the salt screening parameter is decreased by only a factor of 2. 

\subsection{Charge inversion and overcharging}
 
The preceding results suggest that the charge inversion and/or overcharging of the surface,  which are known to occur with asymmetric Coulomb fluids (see, e.g., Refs. \cite{Lemay1,Lemay2,Lemay3,overcharge1,Kjellander,overcharge2,overcharge3,overcharge4,Lozada-Cassou_review,Wang_Gao,Ma,Messina-Holm,Jiang,Shklovs02,SCdressed2} and references therein), may be enhanced when the surface carries a random charge component. This can be inferred from the averaged  cumulative charge, $Q(z)$, being  the sum of the average charges due to fixed and mobile charges (including both monovalent and multivalent ions) within a finite distance $ z$ from the surface, i.e.,
\begin{equation}
Q(z)=\frac {1}{\sigma}\! \int_0^z\!\!\rmd z'\left[e_0n_+(z')-e_0n_-(z')+ e_0q c(z')+\rho_0(z')\right],
\label{eq:Q_av}
\end{equation}
where $n_\pm(z)$ represent the averaged (DH) density of monovalent ions and $c(z)$ the density of dressed multivalent counterions.
As shown in Ref. \cite{SCdressed2}, the cumulative charge can be written only in terms of the counterion density by using the fact for monovalent ions we have $n_+(\Av r)-n_-(\Av r)\simeq -n_b(\beta e_0\psi(\Av r))$, where $\psi(\Av r)$ is the  mean electrostatic potential generated by the explicit charge densities, i.e., 
\begin{equation}
\psi(\Av r)=\int \rmd {\mathbf r}'\, G(\Av r, \Av r')[e_0 q c(\Av r')+\rho_0(\Av r')].
\label{eq:psi_av}
\end{equation}
Hence, using the Green's function expressions  (\ref{eq:inf_green})  for an infinite slab, we have \cite{SCdressed2}
\begin{eqnarray}
Q(\tilde z)&=&-\rme^{-\tilde\kappa \tilde z}+\frac 18 \tilde\chi_c^2  
\label{eq:Q}\\
&\times&\int_0^\infty\!\rmd \tilde z'\left[\textrm{sgn}(\tilde z-\tilde z')\,\rme^{-\tilde \kappa\vert \tilde z-\tilde z'\vert}+\rme^{-\tilde \kappa(\tilde z+\tilde z')}\right]\hat c(\tilde z'), \nonumber
\end{eqnarray}
where   $\hat c(z)\equiv c(z)/c_0$ and $\tilde \chi_c = \chi_c\mu$ with the definition 
\begin{equation}
\chi_c^2 = 8\pi q^2 \ell_{\mathrm{B}} c_0.
\end{equation}

\begin{figure}[t!]\begin{center}
	\begin{minipage}[b]{0.44\textwidth}\begin{center}
		\includegraphics[width=\textwidth]{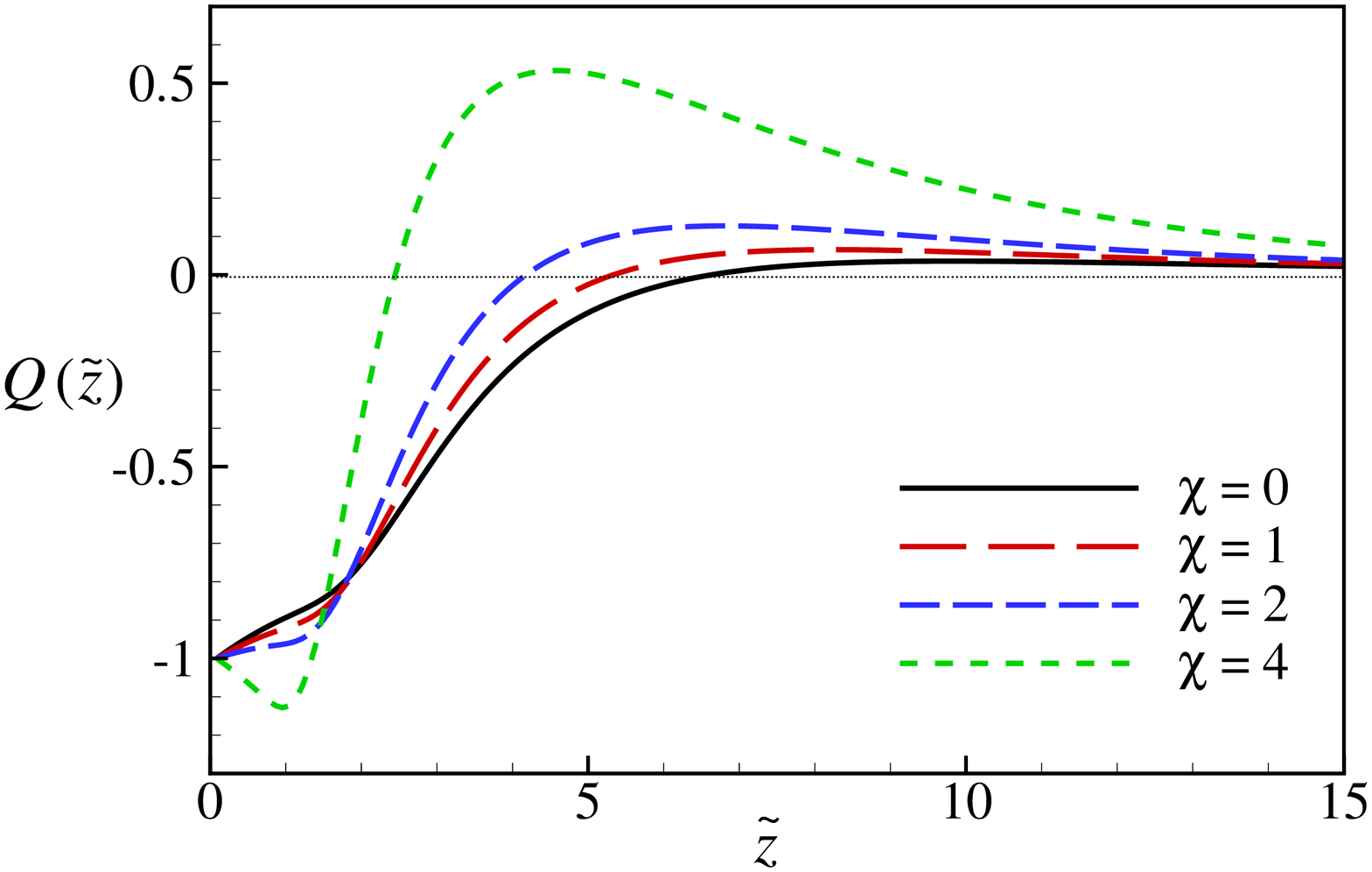} (a) 
	\end{center}\end{minipage}\vskip.3cm
	\begin{minipage}[b]{0.44\textwidth}\begin{center}
		\includegraphics[width=\textwidth]{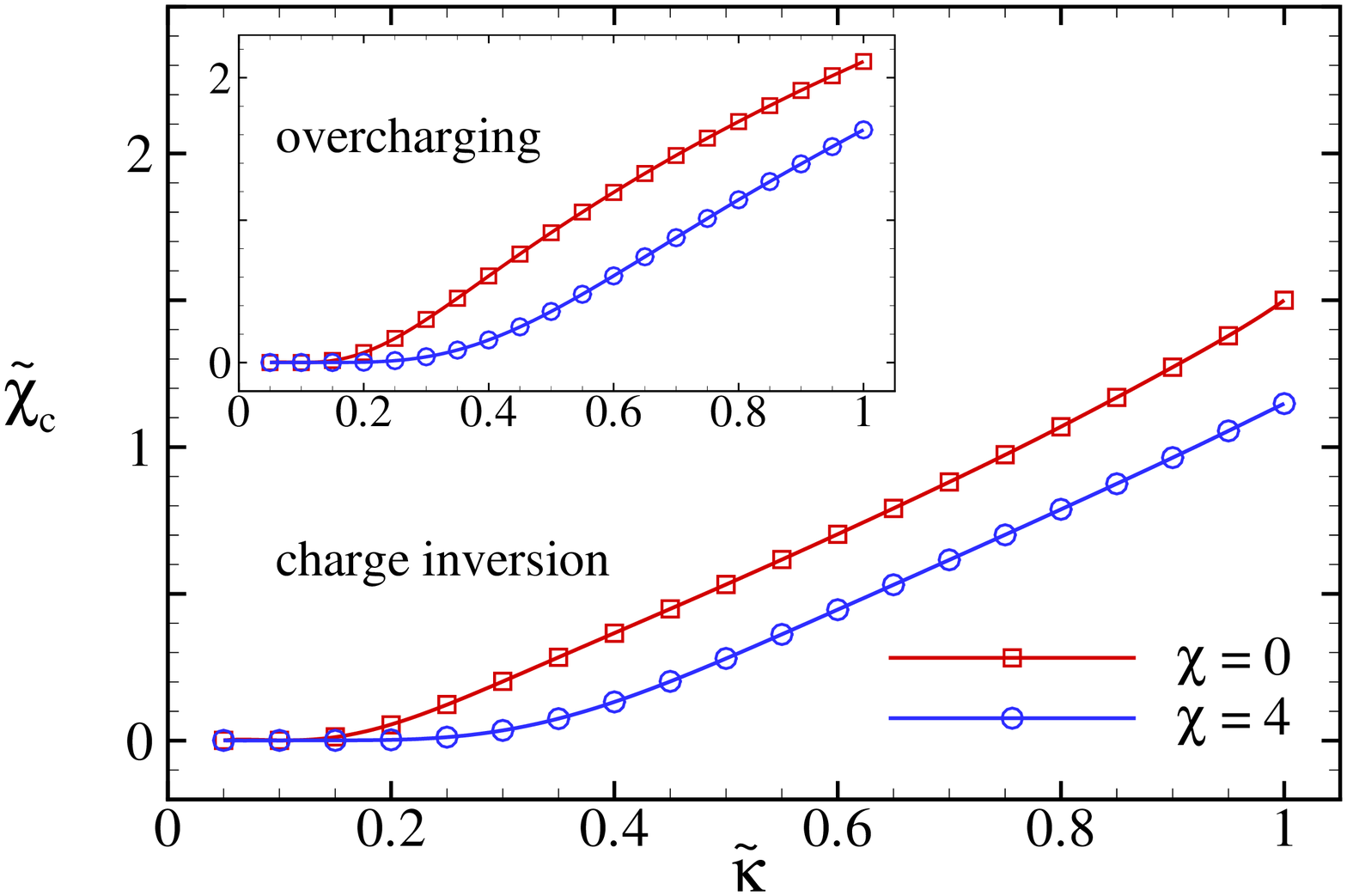} (b)
	\end{center}\end{minipage}
\caption{(Color online)  (a) Averaged cumulative charge next to the randomly charged surface of a semi-infinite slab for $\Xi=50$, $\tilde \kappa = 0.25$, $\tilde \chi_c=0.25$, $\Delta=0.95$  and different values of the  disorder coupling parameter as indicated on the graph.  (b) ``Phase diagram" showing the  minimal amount of multivalent counterion concentration, $\tilde \chi_c$ (in rescaled units), required  to find the charge-inversion (main set) or overcharging  effect (inset) as a function of the rescaled salt screening parameter  in the absence ($\chi=0$) and in the presence of surface charge disorder ($\chi=4$).  The region above the curves shows where   charge inversion (main set) or overcharging (inset)  is predicted to occur.  }
\label{f:phasediag}
\end{center}
\end{figure}

As seen in Fig. \ref{f:phasediag}a, the cumulative charge shows a {\em positive} hump at intermediate separations from the surface, also known as the {\em charge-inversion effect}. The charge-inversion degree  is usually found to amount to a fraction of the total charge (i.e., the maximum value of $Q(\tilde z)$ is smaller than unity)   \cite{SCdressed2,overcharge1,overcharge2,overcharge3,overcharge4,Lozada-Cassou_review,Wang_Gao,Ma,Messina-Holm}. In the presence of charge disorder, a substantially larger amount of multivalent counterions are attracted towards the surface and also a much larger charge-inversion degree (corresponding to the height of the hump) is predicted to occur.

There is a narrow region at small separations from the surface where one can see a decrease in $Q(\tilde z)$ as the disorder strength, $\chi$, is increased. For sufficiently large $\chi$, it exhibits a short-distance dip with $Q(\tilde z)<-1$. In this region, the cumulative charge has the same sign as the {\em mean} surface charge but with a larger magnitude and, therefore, represents the so-called {\em overcharging effect}.  Note that the overcharging effect can be present even in the absence of disorder and depends   on the bulk concentration of multivalent counterions, which enters in Eq. (\ref{eq:Q}) through the rescaled parameter $\tilde \chi_c$. 

In Fig. \ref{f:phasediag}b, we show the minimal amount of bulk multivalent counterion concentration (represented by $\tilde \chi_c$) that is  required  to achieve the charge-inversion (main set)  and overcharging effects (inset) for a wide range of rescaled salt screening parameters. The region above the curves pertains to the parameter values where we find charge inversion or overcharging of the mean surface charge. As seen, for larger salt screening parameters, a larger bulk concentration of multivalent counterions are required to achieve these effects, and for a given salt screening parameter, a larger concentration of multivalent counterions are required to cause overcharging of the surface than its charge inversion. The presence of surface charge disorder facilitates both these effects as they can occur for smaller threshold values of $\tilde \chi_c$, especially at intermediate to large values of the salt screening parameter. 

The aforementioned features of the averaged cumulative charge depend strongly on the dielectric discontinuity at the surface, which generates the image-charge repulsion that competes with the disorder-induced attraction of ions towards the surface. Note that these two latter mechanisms affect  the spatial distribution of multivalent counterions (Figs. \ref{fig:density_inf_delta}a and b), and to a lesser extent, also the spatial distribution of monovalent ions (not shown)   that are treated implicitly in this work; this latter quantity can be calculated from the mean potential, Eq. (\ref{eq:psi_av}), which depends on various system parameters through  the Green's function and the distributions of explicit charges (see Refs. \cite{Ma,SCdressed2,SCdressed3} for explicit-ion simulations of charge inversion and overcharging phenomena at uniformly charged surfaces that incorporate the dielectric image charges as well). The exact form of $Q(\tilde z)$, and the ensuing charge-inversion and/or overcharging effects, thus follow from the interplay between the contributions from the explicit multivalent counterions and the implicit monovalent ions to the averaged cumulative charge at any given set of values for the system parameters. 

Finally, it should be noted that, while the predicted boundaries of the parameter space pertaining to the onset of the charge inversion and/or overcharging are expected to be  relatively accurate \cite{SCdressed2}, the single-particle approximation that lies at the heart of the dressed multivalent-ion description is not expected to be adequate in the  regime of parameters deeply within the regime of charge inversion and/or overcharging due to non-negligible many-body contributions \cite{SCdressed2}.  These considerations and the role of other possible factors such as ion-ion excluded-volume repulsions \cite{overcharge1,overcharge2,overcharge3,overcharge4,Lozada-Cassou_review,Wang_Gao,Ma,Messina-Holm,Jiang,Kjellander}  remain to be assessed further in future simulations.

\begin{figure}[t!]\begin{center}
	\begin{minipage}[b]{0.44\textwidth}\begin{center}
		\includegraphics[width=\textwidth]{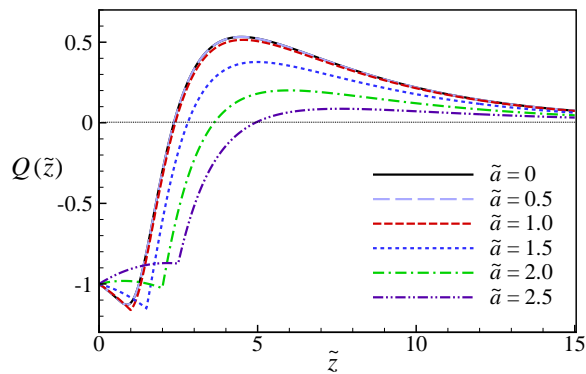} 
	\end{center}\end{minipage}
\caption{(Color online) Averaged cumulative charge next to the randomly charged surface of a semi-infinite slab for $\Xi=50$, $\tilde \kappa = 0.25$, $\tilde \chi_c=0.25$, $\Delta=0.95$, $\chi=4$   and different values of the rescaled radius of the multivalent counterions, $\tilde a=a/\mu$,  as indicated on the graph.}
\label{f:Qc_a}
\end{center}
\end{figure}

\subsection{Counterion size effects}
\label{subsec:ion-size}

We have so far assumed that the multivalent counterions are modeled as point-like particles and can therefore access the whole volume in the region $ z\geq 0$.  On the microscopic level and within the primitive model considered here, the counterion size effects enter through the excluded-volume repulsions between the counterions themselves as well as through the excluded-volume repulsion between individual counterions and the randomly charged surface, which leads to the steric depletion of counterions from the immediate vicinity of the surface. 
Within the strong-coupling approach or its generalization, the dressed multivalent-ion theory on which our approach is based, the partition function is virial expanded systematically to the leading order (Section \ref{sec:formal}) and, as such, involves only the single-particle contributions. This means that only the counterion-surface excluded-volume repulsion enters in the leading-order theory and the excluded-volume repulsions between counterions  enter in the subleading terms that become important at lower electrostatic couplings, falling  outside the regime of interest in the present analysis  \cite{Netz01,AndrePRL,AndreEPJE,hoda_review,perspective,Naji_PhysicaA,asim}. (The dressed multivalent-ion theory of course has a range of validity \cite{SCdressed1,SCdressed2,SCdressed3,perspective} that can be ascertained in detail once simulations or, indeed, alternative theoretical approaches for the present problem become available; see the Discussion). The incorporation of the counterion size effects on the leading virial level is thus straightforward. 

For the sake of simplicity, we model the multivalent counterions as hard-sphere particles with radius $a$. Therefore, one  needs to restrict the volume accessible to multivalent counterions to the region $ z\geq a$, which can be done by re-defining  the indicator function as $\Omega_c({\mathbf r}) = \theta(z-a)$. It is thus evident that  the form of the single-particle interaction energy and, therefore, the $z$-dependent form of the density profile remain unchanged, with the proviso that now one needs to impose the constraint $z\geq a$. The prefactor $\lambda_c$ in Eq. (\ref{eq:sc_density_av}) is still given by the bulk concentration of multivalent counterions, $c_0$, in the presence of a salt bath (Sections \ref{subsec:salt} and \ref{subsec:delta}) and is determined from Eq. (\ref{eq:lambda_c}) within the counterion-only model (Section \ref{subsec:ci-only}). In this latter case, we find the rescaled density profile of hard-sphere counterions of rescaled radius $\tilde a = a/\mu$ as 
 \begin{equation}
\tilde c(\tilde z)\equiv\frac{c(\tilde z)}{2\pi \ell_{\mathrm{B}} \sigma^2}=\frac{\tilde z^{-\frac{\chi}{2}}\;\rme^{-\tilde z}}{\Gamma(1-\frac{\chi}{2}, {\tilde a})}, \qquad \tilde z\geq \tilde a.
\label{eq:dens_a}
\end{equation}
In other words, the singular behavior of the single-particle  interaction energy remains intact and, therefore, the $z$-dependent form of the density profile  still involves an algebraic factor and thus differs qualitatively  from that in the case of a uniformly charged surface, only that the multivalent counterions cannot probe  the singular point at $z=0$ and do not exhibit the diverging surface density obtained with point-like counterions.

The exclusion of multivalent counterions from the vicinity of the surface for large counterion radii can suppress the charge-inversion and/or overcharging effects discussed in the previous Section. In Fig. \ref{f:Qc_a},  we show the averaged cumulative charge, $Q(\tilde z)$, as a function of $\tilde z$ for a few different values of the rescaled counterion radius,   ranging from $\tilde a =0$ up to $2.5$ (in actual units, and assuming $\mu=0.23$~nm, see Table \ref{table}, these values correspond to counterion radii of up to  around 5.8\AA). As seen, upon increasing the counterion radius, the height of the positive hump (the charge-inversion degree) is decreased  and its location shifts to larger distances from the surface. The location of the negative (overcharging) dip also shifts to larger separations; however, the overcharging effect is enhanced at first when the counterion radius is increased and diminishes only when the latter is increased beyond $\tilde a\simeq 1.5$. It is also important to note that for sufficiently small counterion radii, the cumulative charge nearly coincides with that of point-like counterions (black solid curve). In fact, the effects of finite counterion size show up only when the counterion radius becomes larger than the size of the depletion zone generated by dielectric images in the vicinity of the surface, which, in the case Fig. \ref{f:Qc_a}, corresponds to counterion radii $\tilde a \gtrsim 1$ (see also Figs. \ref{fig:density_inf_delta}a and b). In other words, the  charge inversion and/or overcharging of the surface  are affected also by the interplay between the image-charge depletion and the steric depletion of multivalent counterions from the dielectric interface.  

Finally, we note that multivalents counterions may have an internal structure that can introduce higher-order multipolar effects; these effects can be relevant especially for  multivalent counterions with  an extended structure (such as rod-like polyamines including  the trivalent spermidine and tetravalent spermine \cite{spermidine_spermine}), in which case the present hard-sphere model may not be appropriate in order to account for the counterion size effects (see, e.g., Refs. \cite{multipoles,perspective} and references therein).

\begin{table*}[t!]
    \begin{center}
\begin{tabular}{rrrrrrrr|rrrrrrr}
\hline\hline
\multicolumn{8}{c|}{$\Xi=50$\rule{0pt}{3ex}}&\multicolumn{7}{c}{}\\
$\tilde\kappa=0.2$\rule{0pt}{3ex}    &$0.25$    &    $0.3$ &   $0.4$&&&&& &$\chi$&&&$g\,(e_0^2/{\mathrm{nm}}^2$)&& \\
\hline
 \hspace{1ex} $n_0=69$\,mM   &\hspace{1ex} 109\,mM &\hspace{1ex} 158\,mM     &\hspace{1ex} 283\,mM     &\hspace{1ex}$c_0=$1.1\,mM    &\hspace{1ex} $\tilde\chi_c$= 0.1&&&&0.5 &&&0.01&&\\
  57\,mM  & 98\,mM &147\,mM & 272\,mM & 7\,mM &0.25& &&&1.0&&&0.02&&\\
 51\,mM & 91\,mM &140\,mM    &    270\,mM     & 10\,mM& 0.3& &&&2.0&&&0.04&&\\
36\,mM & 75\,mM & 125\,mM &  250\,mM & 18\,mM&0.4& &&&4.0&&&0.08&&\\
\hline\hline
\end{tabular}
\caption{Examples of the actual values of bulk concentrations $n_0$ and $c_0$ that may correspond  to  the typical values of the  rescaled parameters $\tilde\kappa$ and $\tilde\chi_c$ used to plot the figures in the previous Sections. We have chosen $q=4$ (tetravalent counterions), $\ell_{\mathrm{B}}=0.7$~nm and surface charge density $\sigma=0.25\,e_0/$nm$^2$, which give $\mu=0.23$~nm and $\Xi=50$. We also show actual values of the disorder variance $g$ corresponding to a few different  values of the disorder coupling parameter $\chi$  (see the text for definitions).}
\label{table}
\end{center}
\end{table*}

\section{Conclusions and discussion}
\label{sec:discussion}

We have investigated the distribution of multivalent counterions close to a dielectric slab bearing a quenched, random distribution of monopolar surface charges on one of its solution interfaces with set mean value and variance, both in the absence and in the presence of an asymmetric Coulomb fluid, comprised of a mixture of multivalent counterions in a bathing solution of  monovalent ions. Such asymmetric Coulomb fluids are commonplace in many experimental examples such as in the condensation of DNA by multivalent cations in the bulk \cite{Bloom2,Yoshikawa1,Yoshikawa2,Pelta,Plum,Raspaud,Pelta2} 
or in viruses and virus-like nano-capsids \cite{Savithri1987,deFrutos2005,Siber}. Our analysis is done within the framework of the {\em dressed multivalent-ion theory}, which reproduces the strong-coupling theory of multivalent counterions \cite{Netz01,AndrePRL,AndreEPJE,hoda_review,perspective,Naji_PhysicaA,asim} in the zero salt limit and takes into account the  surface-counterion as well as counterion-image correlations on the leading order and in the presence of a bathing salt solution as discussed in detail elsewhere \cite{SCdressed1,SCdressed2,SCdressed3,perspective}. (Note that in the opposite regime of weak coupling, where, e.g., all ions are monovalent, the quenched charge disorder effects turn out to be small and do not lead to any qualitatively new features in the behavior of the system \cite{ali-rudi,netz-disorder,netz-disorder2}.)

In the case of counterions only, we show that a randomly charged surface generates a singular density profile for multivalent counterions with an algebraically diverging behavior at the surface; the latter is characterized by an exponent which is determined by the disorder strength (variance). Thus, multivalent counterions are predicted to accumulate strongly in the immediate vicinity of the randomly charged surface in a way that violates the contact-value theorem, which describes the behavior of counterions at uniformly charged surfaces and predicts a finite contact density \cite{contact_value,contact_value2,contact_value1}. 
This behavior stems from the interplay between the translational entropy of the solution ions and the (non-thermal) configurational entropy due to the averaging over different realizations of the quenched disorder.  Therefore, by introducing an external (quenched) disorder component, we find that the system is driven towards a more `ordered' state characterized by a diminished intrinsic thermal `disorder' in the system. It thus seems appropriate to characterize this response of the system to an externally imposed quenched disorder as the {\em anti-fragile behavior} \cite{taleb} of multivalent counterions in the presence of quenched charge disorder.  It is to be noted that, in the presence of disorder, the system  also attains a thermodynamically more stable state because the internal energy of the system drops in a way that leads to a lowered free energy. 

The singular behavior of multivalent counterions persists also when counterions are immersed into a bath of a monovalent salt solution and there are no dielectric inhomogeneities in the system.  In this case, the slab defines an ion-excluded region, creating salt image effects.
The interplay between the disorder-induced attraction and the salt-image depletion leads to a non-monotonic density profile for counterions close to the surface. The amount of multivalent counterions accumulated near the surface is again  enhanced strongly when the surface is randomly charged.  This holds also in the case of a finite discontinuity in the dielectric constant (even though dielectric image charges, unlike salt images, eliminate the singularity and create a  counterion-depleted zone in the immediate vicinity of the charged surface) and/or when the multivalent counterions  have a finite size (that prevents them from probing the singular point of the single-particle interaction energy on the charged surface). The charge disorder can thus make the overcharging and/or charge inversion of the mean surface charge highly pronounced.

Our results are presented in terms of rescaled (dimensionless) parameters such as the rescaled screening parameter and the electrostatic and disorder coupling parameters, which can be mapped to a wide range of values for counterion and salt bulk concentration, mean surface charge density, counterion valency, etc. A few examples of the actual values for these latter quantities (corresponding to the typical values of the rescaled parameters that were used to plot the figures in the previous Sections) are shown in Table \ref{table}. Note that  other sets of actual parameter values than those given in the Table (e.g., using divalent and trivalent counterions) are just as conceivable, as long as they correspond to the same set of dimensionless parameters. 
The typical values of the disorder coupling parameter that we used in our study, e.g., $\chi\simeq 0-4$, correspond to a relatively small degree of charge disorder on the surface
$g\simeq 0- 0.08\,e_0^2/$nm$^2$. Assuming that the disorder originates from a quenched, random distribution of positive and negative  impurity charges, $\pm e_0$, residing on the surface with a surface density of $n_i$, we find $g = n_i e_0^2$ \cite{disorder-PRL}. Therefore, the above-mentioned values of $g$ can be obtained by relatively small densities $n_i\lesssim 0.1/$nm$^2$ of impurity charges as compared with the mean number of surface charges (typically $\sigma/e_0\lesssim 1/$nm$^2$) and can be thus easily realized in actual systems. Hence, we conclude that the effects due to charge randomness, even at such small amounts, can be quite significant! 

We should emphasize  that our results are valid strictly in the case of highly asymmetric Coulomb fluids, where the dressed multivalent-ion approach can be justified \cite{SCdressed1}.  The  dressed multivalent-ion  theory, that was implemented here, follows as a limiting  single-particle theory from the virial expansion of the partition function up to the leading order in the fugacity of multivalent counterions and, as such, is expected to be applicable in two distinct regimes \cite{SCdressed1,SCdressed2,SCdressed3,perspective}: (i) when the  electrostatic interactions are strong enough giving rise, on the leading order, to a strong-coupling, single-particle behavior for  multivalent counterions next to an oppositely charged boundary  (typically at low salt concentrations or in counterion-only systems) \cite{Netz01,AndrePRL,AndreEPJE,hoda_review,perspective,Naji_PhysicaA,asim}, and (ii) when multi-particle interactions between  counterions are sufficiently weak due, e.g., to high salt screening effects, allowing again for a single-particle description (typically at moderate to high salt concentrations) \cite{SCdressed1,SCdressed2,SCdressed3,perspective}. This analytical approach is thus expected to be valid only at relatively small bulk concentrations of multivalent counterions around,  for instance, just a few mM, which is in fact often the case in  experiments (see, e.g., Refs. \cite{rau-1,rau-2,Bloom2,Pelta,Plum,Raspaud,Pelta2,Yoshikawa1,Yoshikawa2}). 

The dressed multivalent-ion theory has been tested extensively against implicit- and explicit-ion simulations  \cite{SCdressed1,SCdressed2,SCdressed3,perspective,leili1,leili2} and turns out to have a wide range of validity in the parameter space when the surfaces bear uniform charge distributions.  Similar simulations are still missing in the case of randomly charged surfaces with multivalent ions  mostly because of a significantly large increase in the computational time,
which would be required in oder to produce reliable quenched disorder averages. Our results, however, produce concrete predictions that can be tested against simulations. The fingerprints of charge disorder are expected to show up in appropriately designed experiments as well \cite{disorder-PRL,jcp2010,pre2011,epje2012,jcp2012,book,pra2009,pre2010,barrett,speake,kim1,kim2,kim3,kim4,tang}, although one should note that experiments on systems containing solutions of multivalent ions  face  certain difficulties as is, for instance, the case  \cite{Lozada-Cassou_review} in electrophoresis measurements conducted to show the charge inversion effect (see Refs. \cite{Lemay1,Lemay2,Lemay3,Mugele} for other recent methods such as streaming currents or atomic force microscopy measurements). 
In general, we expect that the previously determined regimes of validity of the dressed multivalent-ion theory \cite{SCdressed2} roughly hold also for the present case with disordered surfaces. One particular case that should be treated with caution in systems containing added monovalent salt is the situation where the mean electrostatic potential near the randomly charged surface becomes large, e.g., when the disorder strength is very large and/or the dielectric discontinuity parameter is small, in which case the validity of the underlying DH approximation used for the monovalent ions can break down \cite{SCdressed2}. Another case that goes beyond the present approach is the situation where  nonlinear charge renormalization and/or Bjerrum pairing effects become relevant (see, e.g., Refs. \cite{Kjellander07,Alexander,Bjerrumpairing1,Bjerrumpairing2,Bjerrumpairing3}); however, these effects turn out to be negligible in the regime of parameters that is of concern here \cite{SCdressed1,SCdressed2,SCdressed3,perspective}. 
Also, while we expect that the predicted boundaries of the parameter space pertaining to the onset of the charge inversion and/or overcharging would be relatively accurate, the single-particle approximation that lies at the heart of the dressed multivalent-ion description is not expected to be adequate in the  regime of parameters deeply within the regime of charge inversion and/or overcharging due to non-negligible many-body contributions \cite{SCdressed2}.  These considerations and the role of other possible factors such as higher-order virial corrections \cite{Netz01,AndrePRL,AndreEPJE,hoda_review,perspective,Naji_PhysicaA,asim}, the discrete nature and the finite size of monovalent salt ions \cite{SCdressed2} and the ion-ion excluded-volume repulsions \cite{overcharge1,overcharge2,overcharge3,overcharge4,Lozada-Cassou_review,Wang_Gao,Ma,Messina-Holm,Jiang,Kjellander}, etc, that are expected to become relevant especially at intermediate electrostatic couplings and/or within the regime of charge inversion/overcharging,  remain to be assessed further in future simulations. 

Our model is based on a few simplifying assumptions and, as such, neglects several other factors including  solvent structure (see, e.g., Refs. \cite{Israelachvili,holm,benyaakov,Burak_solvent,Burak_solvent2,Ben-Yaakov2011,Ben-Yaakov2011b}  and references therein), 
 the polarizability  of mobile ions (see, e.g., Refs. \cite{Bikerman,Demery, Hatlo,Netz-polar,Horinek}  and references therein),  specific surface ion-adsorption effects \cite{Belloni95,Forsman06}, etc. We have also neglected the internal structure of counterions that can introduce higher-order multipolar effects (see, e.g.,  Refs. \cite{multipoles,perspective} and references therein); these effects can be relevant especially for  multivalent counterions that have an extended structure such as rod-like polyamines like the trivalent spermidine and tetravalent spermine with chain lengths of up to 1-1.5~nm \cite{spermidine_spermine}. 
 
On the other hand, we have assumed that the charge disorder is distributed according to a Gaussian weight and that it is uncorrelated in space. Spatial correlations can be included in our formalism in a straightforward manner \cite{jcp2010,jcp2012} and will be considered in future works. It is important to note that the precise statistical characteristics of charge disorder in real systems can be highly sample and material dependent, involving also the method of preparation, features that should all be considered if the theoretical findings are to be compared with experiments. 
Furthermore, annealed as opposed to quenched disordered surfaces, containing mobile  surface charges that  are in thermal equilibrium with the  rest of the system  \cite{netz-disorder,netz-disorder2,andelman-disorder,disorder-PRL,jcp2010,safran1,safran2,safran3,Olvera0}, as well as surfaces containing partially quenched or partially annealed charge distributions \cite{partial,Hribar}, and also charge regulating surfaces \cite{Regulation,Regulation2,Regulation3,Olvera,Olvera2,Olvera0,natasha} constitute other interesting examples that can be studied in the present context. All of these additional features we plan to address in the future.

Another interesting problem, which is closely related to the present work and can be studied using similar methods, is the strong-coupling interaction between randomly charged surfaces  immersed in an asymmetric Coulomb fluid \cite{preprint2}. It is also worth mentioning  that  some of the key findings in the present study, such as the  singular behavior of the density profile of multivalent ions,  remain valid even in the case of  {\em net-neutral} surfaces that carry no mean charge density but only a finite charge disorder variance. The case of net-neutral surfaces has been studied recently in a series of works in the context of  Casimir interactions  \cite{disorder-PRL,jcp2010,pre2011,epje2012,jcp2012,book,pra2009,pre2010} and the role of an asymmetric Coulomb fluid in this case will be discussed elsewhere \cite{preprint}.

\begin{acknowledgements}
 
 A.N. acknowledges partial support from the Royal Society, the Royal Academy of Engineering, and the British Academy (UK). 
 H.K.M. acknowledges  support  from the School of Physics, Institute for Research in Fundamental Sciences (IPM), where she stayed as a short-term visiting researcher during the completion of this work.  R.P. acknowledges support from the ARRS through Grants No. P1-0055 and J1-4297.  We thank M. Kandu\v c and E. Mahgerefteh for useful comments and discussions. 

\end{acknowledgements}

\end{document}